\documentclass[12pt,useAMS,useAASTEX,usenatbib]{aastex}
\usepackage{emulateapj5}

\makeatletter

\newenvironment{apjemufigure}{%
\def\@captype{figure}%
\noindent\begin{minipage}{0.999\linewidth}\begin{center}}
{\end{center}\end{minipage}}
\makeatother

\def\l{{\ell}}
\def\lm{\l,m}
\def\alm{a_{\lm}}
\def\ylm{Y_{\lm}}
\def\dl{{\Delta \l}}
\def\ddeltalm{{d^{\Delta}_{\lm}}}
\def\healpix{H{\sc ealpix }}
\def\glesp{G{\sc lesp }}
\def\wmap{\hbox{\sl WMAP~}}
\def\cobe{\hbox{\sl COBE~}}
\def\planck{{\sl Planck~}}

\def\etal{et al.}

\def\cl{C_{\l}}

\def\const{\rm const}

\def\summ{\sum_{m=-\l}^{\l}}
\def\suml{\sum_{\l=2}^{\lmax}}

\def\Cs{{\bf Cs}}
\def\Si{{\bf Si}}

\def\lmin{\l_{\min}}
\def\lmax{\l_{\max}}

\def\lmin{\l_{\min}}
\def\lmax{\l_{\max}}

\newcommand{\nbi}{{Niels Bohr Institute, Blegdamsvej 17,
DK-2100 Copenhagen, Denmark}}

\newcommand{\asc}{{Astro Space Center of Lebedev Physical Institute,
Profsoyuznaya 84/32, Moscow, Russia}}

\slugcomment{Submitted to the Astrophyscal Journal}

\begin{document}

\title{Correlations from Galactic foregrounds in the 1-year Wilkinson Microwave Anisotropy Probe \wmap data}

\author{
Pavel D. Naselsky\altaffilmark{1},
Igor D. Novikov\altaffilmark{1,2},
Lung-Yih Chiang\altaffilmark{1}
}

\altaffiltext{1}{\nbi}
\altaffiltext{2}{\asc}

\email{naselsky@nbi.dk, novikov@nbi.dk, chiang@nbi.dk}


\begin{abstract}

We study a specific correlation in spherical harmonic multipole domain for cosmic microwave background (CMB) analysis. This group of correlation between $\Delta \l=4n$, $n=1,2 \ldots$ is caused by symmetric signal in the Galactic coordinate system. An estimator targeting such correlation therefore helps remove the localized bright point-like sources in the Galactic plane and the strong diffused component down to the CMB level. We use 3 toy models to illustrate the significance of these correlations and apply this estimator on some derived CMB maps with foreground residuals. In addition, we show that our proposed estimator significantly damp the phase correlations caused by Galactic foregrounds. This investigation provides the understanding of mode correlations caused by Galactic foregrounds, which is useful for paving the way for foreground cleaning methods for the CMB.
\end{abstract}

\section{Introduction}
Separation of the cosmic microwave background (CMB) signal from extragalactic and Galactic foregrounds (GF) is one of the most challenging problems for all the CMB experiments, including the ongoing NASA \wmap and the upcoming ESA \planck mission. The GF produces the major (in amplitude) signal in the raw maps, which is localized at a rather small latitude band $b<30^\circ$. To avoid any contribution of the GF to the derived CMB map, starting from \cobe to \wmap experiments, a set of masks and disjoint regions of the map are in use for extraction of the CMB anisotropy power spectrum \citep{wmap,wmapresults,wmapfg,wmapcl,toh,eilc}.
The question is, what kind of assumption about the properties of the foregrounds should we apply for the data processing and what criteria determines the shape and area of the mask and the model of the foregrounds? To answer these questions we need to know the statistical properties of the GF to determine the strategy of the CMB signal extraction from the observational data sets.

These questions are even more pressing for the CMB polarization. Unlike temperature anisotropies, our knowledge about the polarized foregrounds is still considerably poor. Additionally, we have yet to obtain a reasonable truly {\it whole-sky} CMB anisotropy maps for  statistical analysis, while obtaining a whole-sky polarization map seems to be a more ambitious task. Modeling the properties of the foregrounds thus needs to be done for achieving the main goals of the \planck mission: to the CMB anisotropy and polarization signals for the whole sky with unprecedented angular resolution and sensitivity.

Apart from modeling the foregrounds, \citet{toh} (hereafter TOH) propose the ``blind'' method for separation of the CMB anisotropy from the foreground signal. Their method (see also \cite{te96}) is based on minimizing the variance of the CMB plus foreground signal with multipole-dependent weighting coefficients $w(\l)$ on \wmap K to W bands, using 12 disjoint regions of the sky. It leads to their Foreground Cleaned Map (FCM), which seems to be clean from most foreground contamination, and the Wiener-Filtered Map (WFM), in which the instrumental noise is reduced by Wiener filtration. It also provides an opportunity to derive the maps for combined foregrounds (synchrotron, free-free and dust emissions \ldots etc.). Both FCM and WFM show certain levels of non-Gaussianity \citep{tacng,bershadskii,schwarz}, which can be related to the residuals of the GF \citep{magnetic}. Therefore, we believe that it is imperative to develop and refine the ``blind'' methods for the \planck mission, not only for better foreground separation in the anisotropy maps, but also to pave the way for separating CMB polarization from the foregrounds.

The development of ``blind'' methods for foreground cleaning can be performed in two ways: one  is to clarify the multipole and frequency dependency of various foreground components, including possible spinning dust, for high multipole range and at the \planck High Frequency Instrument (HFI) frequency range. The other requires additional information about morphology of the angular distribution of the foregrounds, including the knowledge about their statistical properties in order to construct realistic high-resolution model of the observable \planck foregrounds. Since the morphology of the CMB and foregrounds is closely related to the phases \citep{c3} of $\alm$ coefficients from spherical harmonic expansion $\Delta T(\theta, \phi)$, this problem can be re-formulated in terms of analysis of phases of the CMB and foregrounds, including their statistical properties \citep{phaseentropy,phasemapping,meanchisquare,coleskuiper,pcm,ndv03,ndv04}. 

In \citet{4n}, it is reported that a major part of the GF produces a specific correlation in spherical harmonic multipole domain at $\dl=4$: between modes $a_{\l,m}$ and $a_{\l+4,m}$. The series of  $4n$-correlation from the GF requires more investigation. This paper is thus devoted to further analysis of the statistical properties of the phases of the \wmap foregrounds for such correlation. We concentrate on the question as to what the reason is for the $4n$ correlation in the \wmap data, and can such correlation help us to determine the properties of the foregrounds, in order to separate them from the CMB anisotropies.

In this paper we develop the idea proposed by \citet{4n} and demonstrate the pronounced symmetry of the GF (in Galactic system of coordinates) is the main cause of the $4n$ correlation. The estimator designed in \citet{4n} to illustrate and tackle such correlation can help us understand GF manifestation in the harmonic domain, leading to the development of ``blind'' method for foreground cleaning. In combination with multi-frequency technique proposed in \citet{te96,toh}, the removal of $4n$ correlation of phases can be easily used as an effective method of determination of the CMB power spectrum without Galactic mask and disjoint regions for the \wmap data. It can serve as a complementary method to the Internal Linear Combination method \citep{wmapfg,eilc} and to the TOH method as well, in order to decrease the contamination of
the GF in the derived maps. Such kind of correlation should be observed by the \planck mission and will help us to understand the properties of the GF in details, as it can play a role as an additional test for the foreground models for the \planck mission.

This paper is organized as follows. In Section 2 we describe the $\ddeltalm$ estimator for $4n$-correlation in the coefficients $\alm$ and its manifestation in the observed signals. In Section 3 we apply the estimator on 3 toy models which mimics Galactic foregrounds to investigate the cause of such correlation. In Section 4 we discuss the connection between the $4n$ correlation and the \wmap foreground symmetry. We also examine the power spectrum of the estimator and the correlations of $\ddeltalm$ estimator in Section 5. The conclusion is in Section 6. 

\section{The $4n$ correlation and its manifestation in the \wmap data}
\subsection{The $\ddeltalm$ estimator}
As is shown in \citet{4n} to illustrate the $4n$-correlation, we recap the estimator taken from the combination of the spherical harmonic coefficients $\alm$,
\begin{equation}
d^{\Delta}_{\lm}=\alm - \frac{|a_{\lm}|}{|a_{\l+\Delta,m}|}a_{\l+\Delta,m},
\label{eq1}
\end{equation}
where $|m| \le \l$, and the coefficients $\alm=|\alm| \exp(i\Phi_{\l,m})$ are defined by the standard way:
\begin{equation}
\Delta T(\theta,\phi)= \sum_{\l=2}^{\lmax}\sum_{m=-\l}^\l |\alm| \exp(i\Phi_{\l,m}) Y_{\lm}(\theta,\phi).
\label{eq2}
\end{equation}
$\Delta T(\theta,\phi)$ is the whole-sky anisotropies at each frequency band,
$\theta,\phi$ are the polar and
azimuthal angles of the polar coordinate system, $\ylm(\theta,\phi)$ are the spherical harmonics, $|\alm|$ and $\Phi_{\lm}$ are the amplitudes (moduli) and phases of ${\l,m}$ harmonics. The superscript $\Delta$ in $\ddeltalm$ characterizes the shift of the $\l$-mode 
in $d^{\Delta}_{\l,m}$ and, following \citet{4n}, we concentrate on the series of correlation for $\Delta=4n$, $n=1,2,3\ldots$. Note that the singal of Galaxy mostly
lies close to $\theta=\pi/2$-plane. The estimator $\ddeltalm$ in form Eq.(\ref{eq1}) is closely related with phases of the multipoles of the $\Delta T(\theta,\phi)$ signal. Taking Eq.(\ref{eq2}) into account, we get
\begin{equation}
d^{\Delta}_{\l,m}=\alm\left[1-e^{i(\Phi_{\l+\Delta,m}-\Phi_{\l,m})}\right].
\label{eq3}
\end{equation}
From Eq.(\ref{eq3}) one can see that, if the phase difference $\Phi_{\l+\Delta,m}-\Phi_{\l,m}\rightarrow 0$, then
\begin{equation}
d^{\Delta}_{\l,m}\simeq a_{\lm}e^{-i\pi/2}\sin(\Phi_{\l+\Delta,m}-\Phi_{\l,m})\rightarrow 0.
\label{eq4}
\end{equation}

If $\Phi_{\l+\Delta,m}-\Phi_{\l,m}\ll \pi/2$, the map synthesized from the $d^{\Delta}_{\l,m}$ estimator is simply a map from the $\alm$ with phases rotated by an angle $\pm \pi/2$ and the amplitudes lessened by a factor $|\sin(\Phi_{\l+\Delta,m}-\Phi_{\l,m})|$, while for non-correlated phases $\Phi_{\l+\Delta,m},\Phi_{\lm}$ we have specific (but known) modulation of the $\alm$ coefficients (see the Appendix).

A non-trivial aspect of $d^{\Delta}_{\l,m}$ estimator is that it significantly decrease the 
brightest part of the Galaxy image in the \wmap K-W maps. In the following analysis we use a particular case $n=1$ so that $\Delta=4$, although it can be demonstrated that for $n=2,3\ldots$ the results of analysis do not change significantly as long as $\Delta \le \l_{\rm noise}$, where $\l_{\rm noise}$ is the multipole number in the spectrum where the instrumental noise starts dominating over the GF signal.

\subsection{The $4n$ correlation in the \wmap data}
In this section we show how the $d^{\Delta}_{\lm}$ estimator transforms the GF image in the 
\wmap K-W maps, taking from the NASA LAMBDA archive \citep{lambda}. In Fig.\ref{fig1} we plot the maps synthesized from the $\ddeltalm$ estimator for \wmap K-W band (for $\Delta=4$ and $\lmax=512$)
\begin{equation}
D(\theta,\phi)=\suml \summ d^\Delta_{\lm} \ylm (\theta,\phi).
\label{eq5}
\end{equation} 
Note that the amplitudes are significantly reduced in each map. It should be emphasized that the $D(\theta,\phi)$ map is not temperature anisotropy map, as the phases are altered. 

Let us discuss some of the properties of the $d^{\Delta}_{\lm}$ estimator, which determine the morphology of the $D(\theta,\phi)$ maps. First of all, from Eq.(\ref{eq1}) one can find that for all $m=0$ modes, the estimator is equivalent to zero if ${\rm sign} (a_{\l,0})={\rm sign} (a_{\l+\Delta,0})$ and it is non-zero (and doubled) if ${\rm sign} (a_{\l,0})=-{\rm sign} (a_{\l+\Delta,0})$. In terms of phase difference in Eq.(\ref{eq4}) this means that for $m=0$ modes $d^{\Delta}_{\lm}$ estimator removes those which have the same phases, while doubles the amplitudes of others whose phases differing by $\pi$ in the $D(\theta,\phi)$ maps.

\begin{apjemufigure}
\centerline{\includegraphics[width=0.65\linewidth]{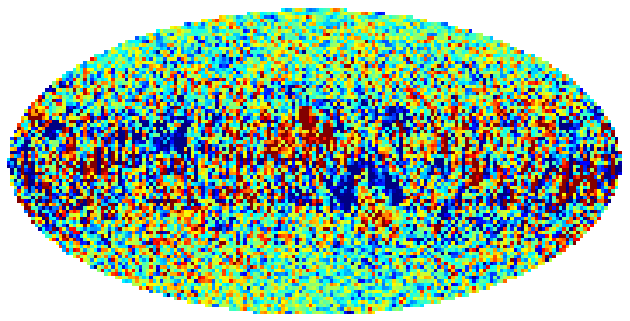}}
\centerline{\includegraphics[width=0.65\linewidth]{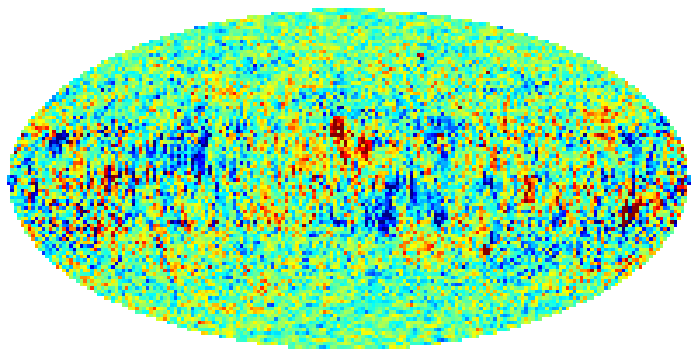}}
\centerline{\includegraphics[width=0.65\linewidth]{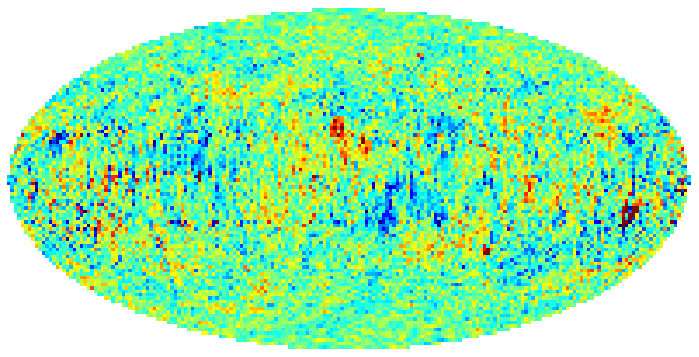}}
\centerline{\includegraphics[width=0.65\linewidth]{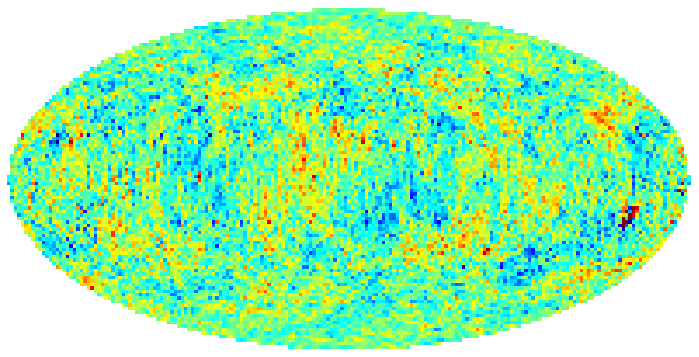}}
\centerline{\includegraphics[width=0.65\linewidth]{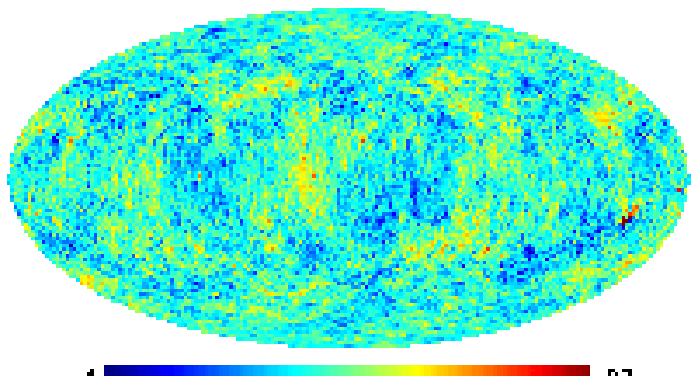}}
\caption{The $D(\theta,\phi)$ maps (from the top to the bottom) synthesized from the $\ddeltalm$ estimator for \wmap K, Ka,Q, V and W maps. Note that the colorbar limits from top to bottom are $[-0.50,0.50]$, $[-0.50,0.50]$, $[-0.50,0.50]$, $[-0.40,0.40]$ and $[-0.38,0.50]$, respectively.}
\label{fig1}
\end{apjemufigure}

However, such specific case of $d^{\Delta}_{\l,m}$ estimator for $m=0$ modes is not unique for
$\Delta=4n$ only. It seems typical for any values of $\Delta$ parameter. What is unique in the \wmap data is that for $\Delta=4 n$ the order of sign for $m=0$ modes leads to the $D(\theta,\phi)$ image without strong signal from the Galactic plane.

We present in Fig.\ref{fig2} the images synthesized of the even and odd $m$ modes from the \wmap W band signal. The even and odd $m$ modes reflect different symmetry of the signal, related to the properties of the spherical harmonics and the corresponding symmetries of the foregrounds. For even $m$ the brightest part of the signal is mainly localized in the Galactic plane area (the top panel), while for odd $m$ modes the signal has less dominated central part from the GF, but it has well presented periodic structure in $\theta$ direction (horizontal stripes), from the north to south pole caps crossing the Galactic plane.
In Fig.\ref{fig3} we present the symmetry of the GF for the W band signal for even and odd $\l$ harmonics, including all corresponding $m$ modes.

\begin{apjemufigure}
\centerline{\includegraphics[width=0.65\linewidth]{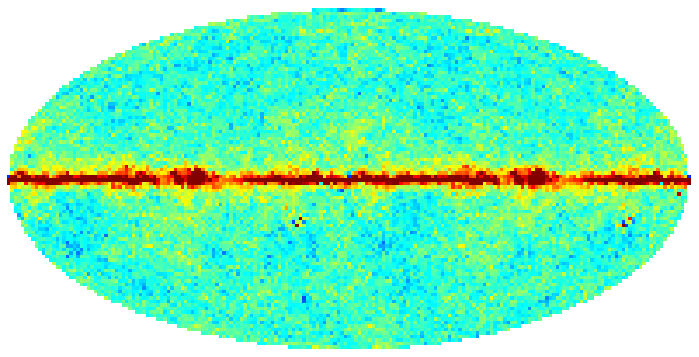}}
\centerline{\includegraphics[width=0.65\linewidth]{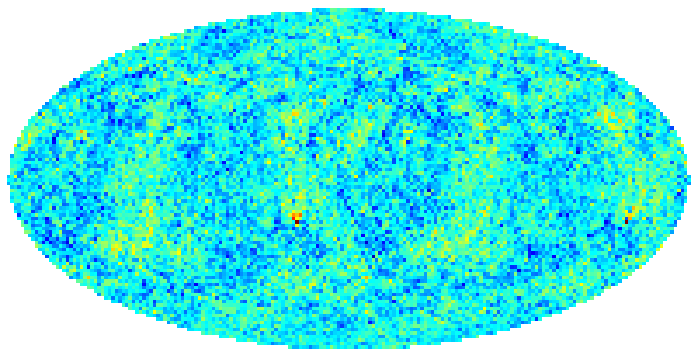}}
\centerline{\includegraphics[width=0.65\linewidth]{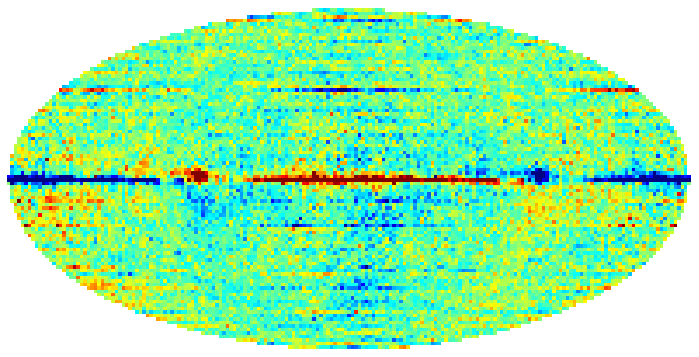}}
\centerline{\includegraphics[width=0.65\linewidth]{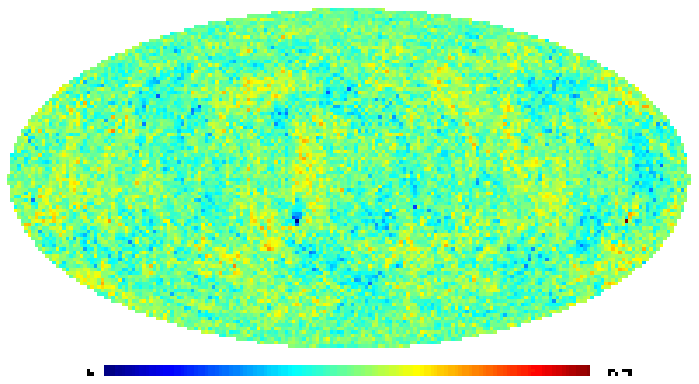}}
\caption{Comparison between even and odd $m$ modes. The top panel is the map synthesized of the even $m$ modes for $2 \le \l \le 512$ from the \wmap W band signal. The 2nd is the $D(\theta,\phi)$ map synthesized from the $d^{\Delta}_{\l,m}$ estimator on the 1st panel. The 3rd is that synthesized of the odd $m$ modes from the \wmap W band signal, and the 4th is the $D(\theta,\phi)$ map on the 3rd panel. All the maps are plotted with colorbar limit  $[-0.5,0.5]$ mK.}
\label{fig2}
\end{apjemufigure}

As one can see from Fig.\ref{fig2}-\ref{fig3}, the even $m$ and the even $\l$ maps (the top of Fig.\ref{fig2} and Fig.\ref{fig3}) have a common symmetrical central part, which looks like a thin belt covered in $\theta \sim \pi/2$ area and all $0 < \phi \le 2\pi$ range. For odd modes the brightest GF mainly concentrate locally in $\theta$ and
$\phi$ rectangular areas. Additionally, for the maps of even and odd $\l$ harmonics in Fig.\ref{fig3} we have periodic structure of the signal in $\theta$ direction, which is determined by the properties of the spherical harmonics, and, more importantly, by the properties of $a_{\l,m}$ coefficients of decomposition, which reflect directly corresponding
symmetry of the GF.

\begin{apjemufigure}
\centerline{\includegraphics[width=0.65\linewidth]{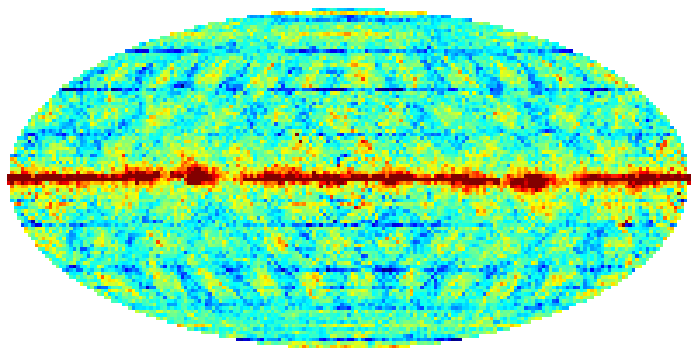}}
\centerline{\includegraphics[width=0.65\linewidth]{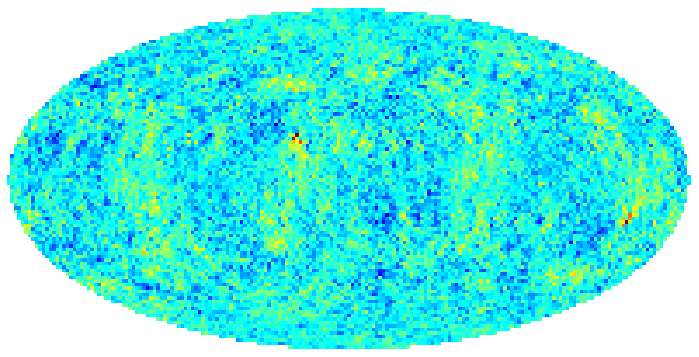}}
\centerline{\includegraphics[width=0.65\linewidth]{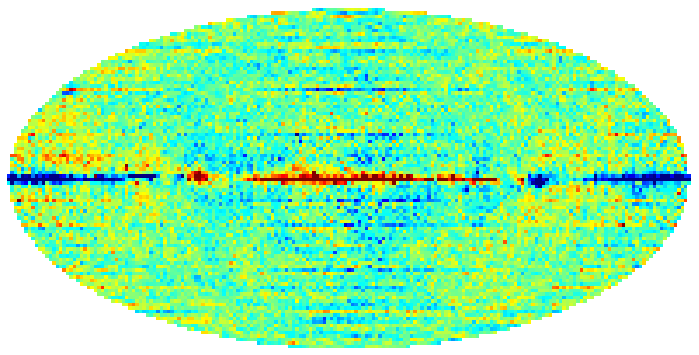}}
\centerline{\includegraphics[width=0.65\linewidth]{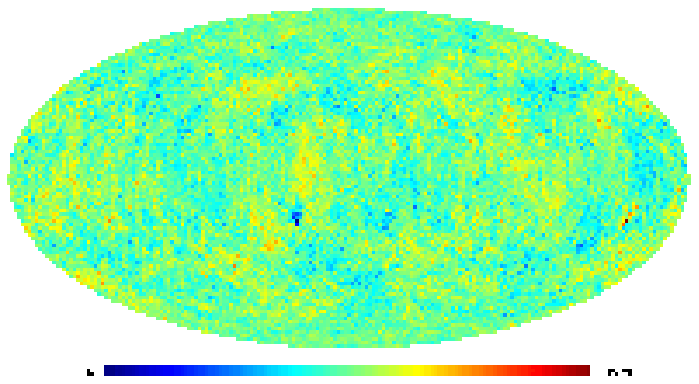}}
\caption{Comparison between even and odd $\l$ modes. The top panel is the map synthesized of the even $\l$ modes for $2\le \l \le 512$ from the \wmap W band signal. The 2nd is the $D(\theta,\phi)$ map synthesized from the $d^{\Delta}_{\l,m}$ estimator on the 1st panel. The third is that synthesized of the odd $\l$ modes from the W band and the 4th the $D(\theta,\phi)$ map on the 3rd panel. All the maps are plotted with colorbar limit  $[-0.5,0.5]$ mK.}
\label{fig3}
\end{apjemufigure}

\section{Why does the 4n correlation appear?}
In this section we want to examine why the $4n$ correlation appears. In order to answer this question we introduce 3 toy models for the Galaxy emissivity, which reflect directly different symmetries of the Galactic signal. In Appendix we analyze more general situation. These 3 toy models are the belt, the rectangular and the spots models. All 3 models are the simple geometrical shapes added with the \wmap ILC map. 

\begin{enumerate}
\item The belt model:

 We add on top of the \wmap ILC map with
\begin{equation}
\Delta T_b(\theta,\phi)=A=\const,
\label{belt}
\end{equation}
if $\pi/2-\delta\le\theta\le\pi/2+\delta$,and $\phi\subseteq [0,2\pi]$, where $\delta$ is the halfwidth of the belt. We set $A=10$ mK and $\delta=5^\circ$.

\item The rectangular model:

We add on top of the \wmap ILC map with
\begin{equation}
\Delta T_r(\theta,\phi)=A=\const
\label{rec}
\end{equation}
for $\pi/2 - \delta \le \theta \le \pi/2 + \delta$, and $\gamma \le \phi \le 2 \pi- \gamma$. We further use two variants
\begin{itemize}
\item $\delta=5^\circ$ and $\gamma=\pi$,
\item $\delta=5^\circ$ and $\gamma=\pi/4$.
\end{itemize}

\item The spots model:

This model is to mimic the properties of symmetric bright point-like sources convolved with a Gaussian beam $B(\theta,\phi)$ with FWHM=$5^\circ$. So on ILC map we add

\begin{eqnarray}
\Delta T_s(\theta,\phi)&=&\sum_j \int A_j \delta^D(\cos\theta_j) \delta^D(\phi -\phi_j) \nonumber \\ 
&& \times B(\theta-\theta_j,\phi -\phi_j) d\theta_j d\phi_j,
\end{eqnarray}
where $\delta^D$ is the Dirac-$\delta$ function, and the amplitudes of the point sources $A_j$ are in order of 10 mK.
\end{enumerate}
Below we examine these toy models to see how they can illustrate the $4n$ correlation. 

\subsection{The belt model}
Ideologically this model best illustrates that the symmetry of the GF signal can help remove the GF itself without any additional assumption about properties of the foregrounds and of Galactic mask as well. The theoretical basis is that the properties of the phases for the belt signal are related to $m=0$ modes (see below). Thus, any method that can remove all  $m=0$ modes from the map is automatically able to remove the belt-like Galactic signal and reproduce the ILC signal for all $m\neq 0$ modes without any errors.
However, these $m=0$ modes contribute significantly to the low multipole part of the power spectrum $\cl$ for the reconstructed CMB signal (which is in our model the ILC signal for $\l,m\neq 0$ modes). Using $d^{\Delta}_{\l,m}$ estimator we can avoid the problem of reconstruction of the $m=0$ modes and the corresponding power spectrum of the $D(\theta,\phi)$ map (see Section \ref{power}).

In Fig.\ref{sim} we plot the resultant $D(\theta,\phi)$ maps for the belt-like model. As one can see, the particular case seems peculiar for the $d^\Delta_{\lm}$ estimator, which does not remove the Galactic signal properly. According to the private communication with Eriksen et al., this simple model of the Galactic signal should reflect the properties of spherical harmonics, namely the $4n$ correlation mentioned in \citet{4n}. The symmetry of this model is related to $2n$ correlation of even harmonics, while for odd harmonics the Galactic signal must vanish.
However, for some of the $m=0$ harmonics the $d^{\Delta}_{\l,m=0}$ estimator differs from zero, and the corresponding stripes determines the morphology of the image (see Fig.\ref{sim}, the second from the top).
To show that effect clearly, in Fig.\ref{sim} (the third) we plot the map, which contains the non-zero $d^{\Delta}_{\l,m=0}$ modes. Remarkably, the morphology of defects in the second  and third one are practically the same. To demonstrate that, in the $D(\theta,\phi)$ map (the second in Fig.\ref{sim}) we simply remove all non-zero $d^{\Delta}_{\l,m=0}$ modes and get the map shown in Fig.\ref{sim} as the
4th. So, the second and the 4th maps are the same, excluding all non-zero modes $d^{\Delta}_{\l,m=0}$ from the $D(\theta,\phi)$ map. This result is not surprising, if one takes into account the symmetry of the belt signal, all the phases of $\l,m$ harmonics for $m\neq 0$ are exactly those of $\l,m\neq 0$ harmonics in the ILC map, while all $a_{\l,m=0}$ in the first from the top in Fig.\ref{sim} map are mainly determined by the belt signal (Galaxy). To show that in Fig.\ref{sim} we simply remove all $a_{\l,m=0}$ modes from the ILC plus the belt map and the resultant is shown in Fig.\ref{sim} the bottom map. One can see, that this is the ILC map for all $a_{\l,m\neq 0}$ harmonics.
What properties of this toy model are crucial in producing the $4n$ multipole correlation of the phases? Taking the definition of the belt signal into account, one can obtain that for the belt signal

\begin{equation}
a^b_{\l,m}=2A\left[\delta_{\l,\l=0}- P^{-1}_\l(\cos\Theta) \sin \Theta \delta_{\l,2n}\delta_{m,m=0}\right],
\label{bel}
\end{equation}
where $n=1,2,\ldots$, $\delta_{x,y}$ is the Kronecker symbol, $P^{-1}_\l(\cos\Theta)$ is the Legendre polynomials and $\Theta=\pi/2-\delta$. For $\l\delta\gg 1$ the asymptotic of $a^b_{\l,m}$ is \citep{gr}
\begin{equation}
a^b_{\l,m=0}\simeq -\frac{4A \Gamma(\l)}{\sqrt{2\pi}\Gamma(\l+\frac{3}{2})} \sin\left[(\l+\frac{1}{2})\delta\right]\delta_{\l,2n}.
\label{bel1}
\end{equation}
Then for $d_{\l,0}$ from Eq.(\ref{eq2}) we obtain
\begin{eqnarray}
d^{\Delta}_{\l,0}&=&-\frac{4A \Gamma(\l)}{\sqrt{2\pi}\Gamma(\l+\frac{3}{2})} \left\{\sin\left[(\l+\frac{1}{2})\delta\right] \right. \nonumber \\
&&\left. -\frac{|\sin\left[(\l+\frac{1}{2})\delta\right]|}
{|\sin\left[(\l+\Delta+\frac{1}{2})\delta\right]|} \sin\left[(\l+\Delta+\frac{1}{2})\delta\right]\right\}.
\label{bel2}
\end{eqnarray}
As is seen from Eq.(\ref{bel2}), $d_{\l,0}$ depends on the sign of the first and second terms in the brackets. Let us assume that for some $\l=\l^{*}$ the first term is positive and the second term is negative. In this case

\begin{eqnarray}
d^{\Delta}_{\l,0}\sim\sin\left[(\l+\frac{1}{2})\delta\right]
\sin\left[(\l+\Delta+\frac{1}{2})\delta\right],
\label{bel3}
\end{eqnarray}
and unlike compensation of the modes $d^{\Delta}_{\l,0}=0$, we have the signal in order of magnitude close to $-4A /\sqrt{2\pi} \l^{-\frac{3}{2}}$.
Moreover, taking into account that the amplitude of that signal decreases very rapidly in comparison with amplitudes of the CMB signal, starting from some multipoles $\l=\l_{cr}\sim \delta^{-1}$ the order of sign in Eq.(\ref{eq2}) is determined by the CMB
signal and the number of horizontal stripes increase rapidly, as it is seen from the Fig.\ref{sim} (the second map from the top).

\begin{apjemufigure}
\centerline{\includegraphics[width=0.65\linewidth]{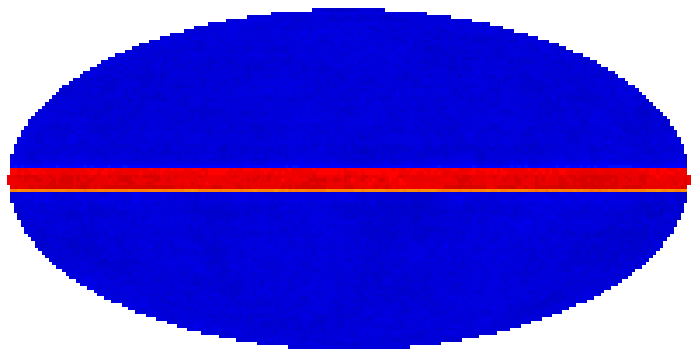}}
\centerline{\includegraphics[width=0.65\linewidth]{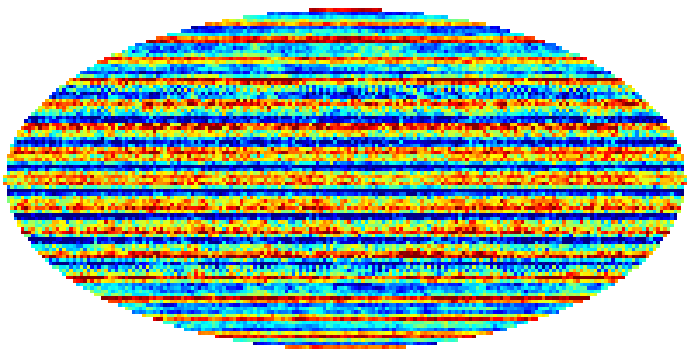}}
\centerline{\includegraphics[width=0.65\linewidth]{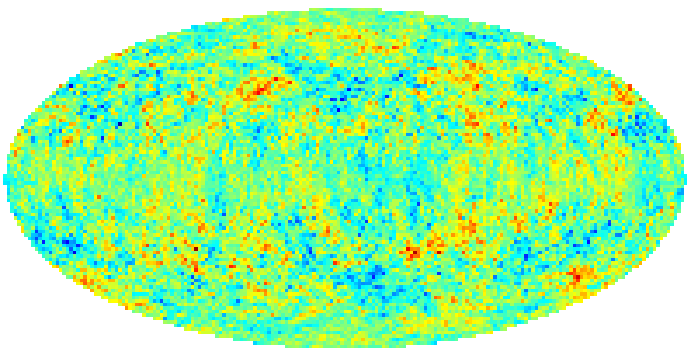}}
\centerline{\includegraphics[width=0.65\linewidth]{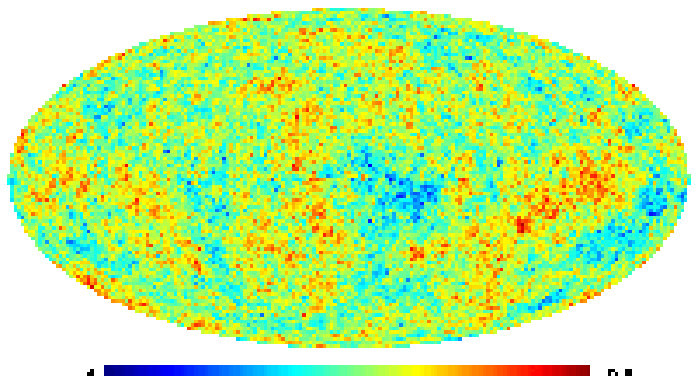}}
\caption{The belt model. The 1st panel is the \wmap ILC map plus artificial uniform Galactic belt signal (see text). The 2nd panel is the $D(\theta,\phi)$ map synthesized from $\ddeltalm$-estimator ($\l \le 250$) on the 1st panel. The 3rd panel is also the $D(\theta,\phi)$ map from $\ddeltalm$-estimator (the same as the 2nd), but with all $m=0$ modes in $\ddeltalm$ being set zero. The bottom panel is synthesized from the $\alm$ of the 1st panel, but with all $m=0$ modes in the $\alm$ being set zero. The colorbar limit for the first panel is $[-0.6,10.5]$mK, and the rest are $[-0.5,0.5]$mK.}
\label{sim}
\end{apjemufigure}

Thus, for a given "belt" model of the Galaxy emissivity for the range of multipoles $\l \delta \gg 1$ the $4n$ multipole correlation does not exist.
However, in the opposite approximation $\l \delta\ll 1$ the properties of the multipole coefficients are determined by the asymptotic of Legendre polynomials in Eq.(\ref{bel})

\begin{eqnarray}
a^b_{\l,m}
\sim\left[\frac{\Gamma(\l+2)}{\Gamma(\l)\Gamma\left(1+\frac{\l+1}{2}\right)\Gamma\left(\frac{2-\l}{2}\right)}\right]^{\frac{1}{2}},
\label{bbel}
\end{eqnarray}
and formally $a^b_{\lm}=0$ for all $\l=2n$, $n=1,2\ldots$,but $\l\delta\ll 1$.
As follows from Eq.(\ref{bbel}), the symmetry of the belt model does not require $4n$ correlation at all. If $\Delta$ parameter is even $\Delta=2n$, the sum $\l+\Delta$ is even too, and the corresponding $a^b_{\l,m}$ coefficient vanishes.
The case for $\Delta=4n$ is a special one for the more general correlation $\Delta=2n$, which will be broken at the limit $\l\delta\gg 1$. However, the belt model as discussed above provides a useful estimation of the properties of the
$\ddeltalm$ estimator for more general cases, when the symmetry of the GF is not so high as in the belt toy model. Particularly, to prevent any contribution to the $D(\theta,\phi)$ map from the highly symmetrical part of the GF, we will use further generalization of the $\ddeltalm$ estimator as follows:
\begin{eqnarray}
d^{\Delta}_{\l,m}&=& a_{\l,m}-\frac{|a_{\l,m}|} {|a_{\l+\Delta,m}|} a_{\l+\Delta,m},\hspace{0.5cm} {\rm if} \hspace{0.3cm} m\neq 0 ; \nonumber\\
d^{\Delta}_{\l,m}&=& 0, \hspace{3.9cm} {\rm if} \hspace{0.3cm} m=0.
\label{eq1b}
\end{eqnarray}
We will use this generalization for $\ddeltalm$ in all the subsequent discussions (excluding descriptions of the toy models in \S 3.2 and 3.3).

\subsection{The rectangular model}
Let us discuss the properties of the model which mimics the Galaxy image in the map as a rectangular area, characterized by halfwidth $\delta$ in $\theta$ direction and $\Phi$ in $\phi$ direction and with a constant amplitude of the signal $A$ in the area $\pi/2-\delta\le \theta\le \pi/2+\delta$, $\Phi\le \phi\le 2\pi -\Phi $. For that model the corresponding
 $a_{\l,m}$ coefficients are
\begin{eqnarray}
a^r_{\l,m}&=&-2A\left[\frac{(2\l+1)\Gamma(\l-m+1)}{4\pi \Gamma(\l+m+1)}\right]^{\frac{1}{2}}\frac{\sin(m\Phi)}{m} \nonumber\\
&\times &\int_{\pi/2-\delta}^{\pi/2+\delta}d\theta\sin\theta P_\l^m(\cos\theta).
\label{rec}
\end{eqnarray}

The properties of the integral in Eq.(\ref{rec}) depend on the parameter $\l\delta$. If $\l\delta\ll 1$, then
$P_\l^m(\cos\theta)\simeq P_\l^m(0)$, where
\begin{equation}
P_\l^m(0)= \sqrt{\pi} 2^m \left[\Gamma\left(1+\frac{\l-m}{2}\right)\Gamma\left(\frac{1-\l-m}{2}\right)\right]^{-\frac{1}{2}},
\label{rec1}
\end{equation}
while for $\l\delta\gg 1$ we get \citep{gr}
\begin{eqnarray}
P^m_\l(\cos\theta_k)\simeq \frac{2}{\sqrt{2\pi\sin\theta_k}}\frac{\Gamma(\l+m+1)}{\Gamma(\l+\frac{3}{2})}\nonumber\\
\times \cos\left[(\l+\frac{1}{2})\theta_k +\frac{1}{2} m\pi -\frac{\pi}{4}\right].
\label{rec2}
\end{eqnarray}
Thus, for these asymptotics we have
\begin{eqnarray}
a^r_{\l,m}
\sim\left[\frac{\Gamma(\l-m+1)}{\Gamma(\l+m+1)\Gamma\left(1+\frac{\l-m}{2}\right)\Gamma\left(\frac{1-\l-m}{2}\right)}\right]^{\frac{1}{2}}
\label{rec3}
\end{eqnarray}
if $\l\delta\ll 1$, and
\begin{eqnarray}
a^r_{\l,m}&\sim&\left[\frac{\Gamma(\l-m+1)}{\Gamma(\l+m+1)}\right]^{\frac{1}{2}}
\left[\frac{\Gamma(\l+m+1)}{\Gamma(\l+\frac{3}{2})}\right] \nonumber\\
&&\times \left\{ \cos(\frac{\pi(\l+m)}{2})\sin \left[(\l+\frac{1}{2})\delta \right]- \right. \nonumber\\
&&\left.\frac{(2\l+1)\delta}{\pi(\l+m)}\sin\left[\frac{\pi(\l+m)}{2}+(\l+\frac{1}{2})\delta\right] \right \} \nonumber\\
\label{rec4}
\end{eqnarray}
if $\l\delta\gg 1$.
For odd $\l+m$, as seen from Eq.(\ref{rec3}), $a^r_{\l,m}=0$ if $\l\delta\ll 1$\footnote{The first non-vanished for odd $\l+m$ is in order of magnitude $(\l\delta)^3$}. That means that the main term in Eq.(\ref{rec}), proportional to $\delta/\pi$ is related with the even harmonics $\l+m=2n, n=1,2 \ldots$. The leading term, which determines the sign of $a^r_{\l+\Delta,m}$ is $\Gamma(\frac{1-\l-m-\Delta}{2})$ in the denominator of Eq.(\ref{rec4}).
For that term we have
\begin{equation}
\Gamma(\frac{1-\l-m-\Delta}{2})=\frac{
\Gamma\left(\frac{1-\l-m}{2}\right)}{\Pi_{j=1}^{\frac{\Delta}{2}}\left(\frac{1-\l-m}{2}-j\right)},
\end{equation}
and for $\Delta=4k, k=1,2\ldots$ the sign of $a^r_{\l+\Delta,m}$ is the same, as for  $a^r_{\l,m}$. Taking into account that for $d^r_{\l,m}$ estimator the order of signs for $\l,m$ and $\l+\Delta,m$ is crucial, we can conclude that the 
compensation of the central part of the signal requires $\Delta=4k, k=1,2\ldots$. However, it does not guarantee that
for $\l\delta\gg 1$ modes $\Delta=4k$ criteria leads the compensation of the signal. To show this let us describe the asymptotic $\l\delta\gg 1$, when the symmetry of the $a_{\lm}$ coefficients is determined by Eq.(\ref{rec2}).
As one can see from this equation, if $\l+m= 4k, k=1,2\ldots$ the sign of $a^r_{\l+\Delta,m}$ is now determined by the combination
$\sin\left[(\l+\frac{1}{2})\delta\right]$ and does not show any $4k$-correlation at all. Moreover, according to the properties of the sine mode the shift of the argument $\l$ by the factor $\Delta=4k$ just transforms it to the combination
\begin{eqnarray}
\sin\left[(\l+4k+\frac{1}{2})\delta\right]&=&\sin\left[(\l+\frac{1}{2})\delta\right]\cos(4k\delta)\nonumber\\
&+&\cos\left[(\l+\frac{1}{2})\delta\right]\sin(4k\delta).
\label{sin}
\end{eqnarray}
\begin{apjemufigure}
\centerline{\includegraphics[width=0.60\linewidth]{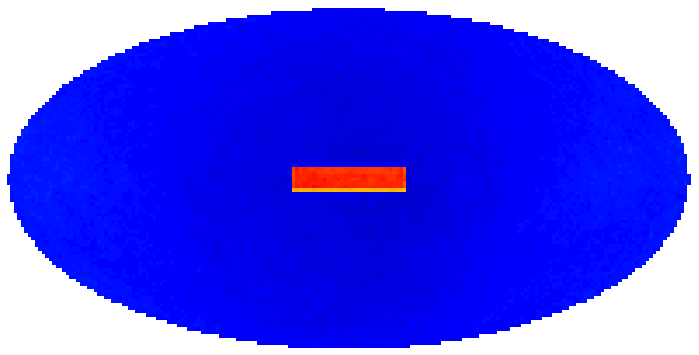}}
\centerline{\includegraphics[width=0.60\linewidth]{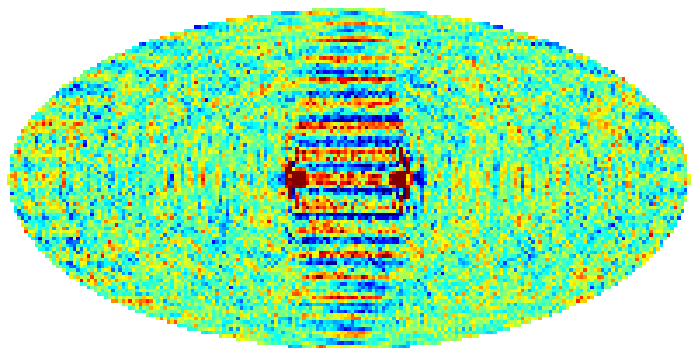}}
\centerline{\includegraphics[width=0.60\linewidth]{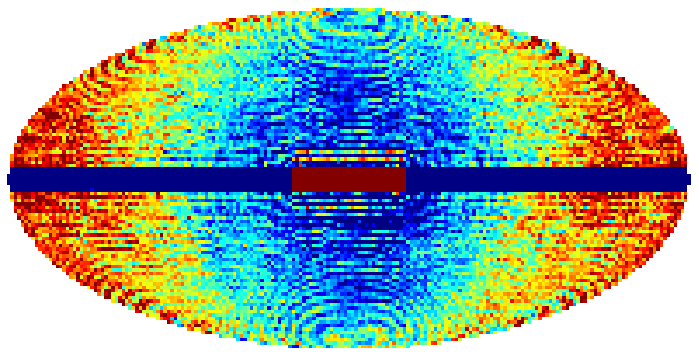}}
\centerline{\includegraphics[width=0.60\linewidth]{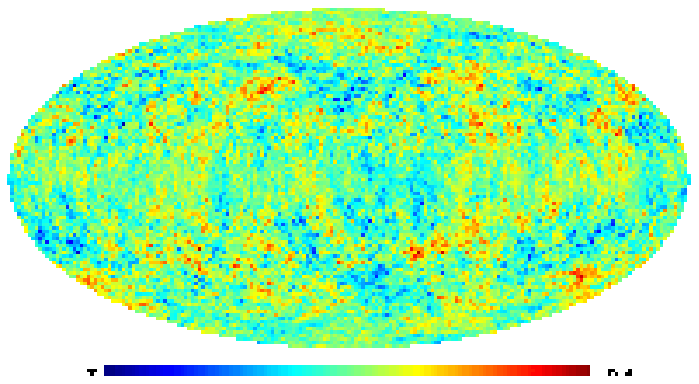}}
\caption{The rectangular model. The 1st panel is the \wmap ILC map plus the rectangular signal (see text). The 2nd is the $D(\theta,\phi)$ map synthesized from the $\ddeltalm$-estimator without Eq.(\ref{eq1b}) generalization ($\l \le 250$) on the 1st panel. The 3rd is from the 1st panel, but with all $m=0$ modes in the $\alm$ being set zero. The bottom is the $D(\theta,\phi)$ map from the $\ddeltalm$-estimator (without Eq.(\ref{eq1b}) generalization) on the ILC map alone. The colorbar limit for the 1st panel is $[-0.6,10.5]$mK, and the rest are $[-0.5,0.5]$mK.}
\label{fign}
\end{apjemufigure}

Thus, one can see that $4k$-correlation requires some restriction on the $\delta$-parameter
\begin{equation}
4k\delta=2\pi m.
\label{cos}
\end{equation}
Thus, for $k=1$ and $m=1$ the halfwidth of the rectangular area must be close to $\delta=\pi/2$. If, for example,
$\delta=\pi/4$ then we will have correlation for $k=2$, but not for $k=1$. Practically speaking, for $\delta \l \gg 1$ we can have some particular symmetry, but not general symmetry $\Delta=4n$. Conclusions concerning this rectangular model of GF are clearly seen in Fig.\ref{fign}.

\subsection{The spot model}
To understand how each sort of defects is related with corresponding $4n$-correlation, we introduce the model of defects, which can be describe as a sum of peaks with amplitudes $A_j$ and coordinates $\theta_j,\phi_j$. For analytical description of the model we neglect the beam convolution of the image of point sources (PS), but we include it in the numerical simulation. For the model of defects
\begin{equation}
\Delta T(\theta,\phi)=\sum_j A_j\delta(\cos\theta -\cos\theta_j)\delta(\phi -\phi_j).
\label{ps}
\end {equation}
For the $\alm$ coefficients of the spherical harmonics expansion from Eq.(\ref{ps}) we get
\begin{equation}
\alm=\sum_j A_j\sqrt{\frac{2\l+1}{4\pi}\frac{(\l-m)!}{(\l+m)!}}P^m_\l(\cos\theta_j)e^{-im\phi_j}.
\label{ps1}
\end {equation}
As for the rectangular model, we will assume that all $\theta_j=\pi/2$ and simply we will have
$a_{\l,m}\sim P_\l^m(0)$ for $\l\delta\gg 1$, as well as $\l\delta\ll 1$. As it was shown in previous section,
$P_\l^m(0)$ clearly demonstrate $4n$ correlation. We would like to point out that for the spots model this correlation now is strong, unlike the belt and rectangular models. \footnote{Strongly speaking, we should use the terms $\l+m=4n$-correlation.} Moreover, implementation of the Gaussian shape of the PS which come from beam convolution does not change that symmetry at all. To show that, in Fig.\ref{ps} we plot the model of two PS with amplitudes in order to
10 mK, combined with the ILC map. The reason for such effect is quite obvious. The beam convolution does not change the symmetry of the model, but rescale the amplitudes of the PS by a factor $\exp[-\l(\l+1)]/2\sigma^2$, if we assume the Gaussian shape of the beam.

\begin{apjemufigure}
\centerline{\includegraphics[width=0.62\linewidth]{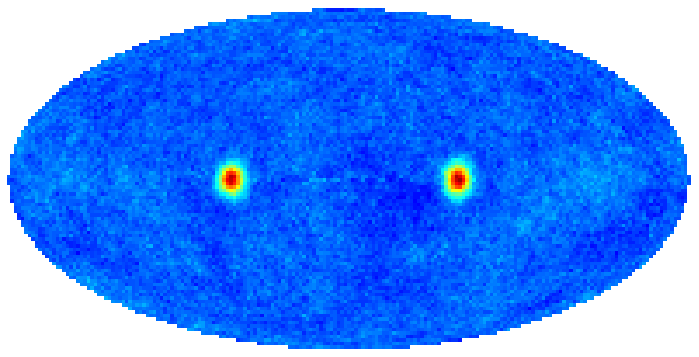}}
\centerline{\includegraphics[width=0.62\linewidth]{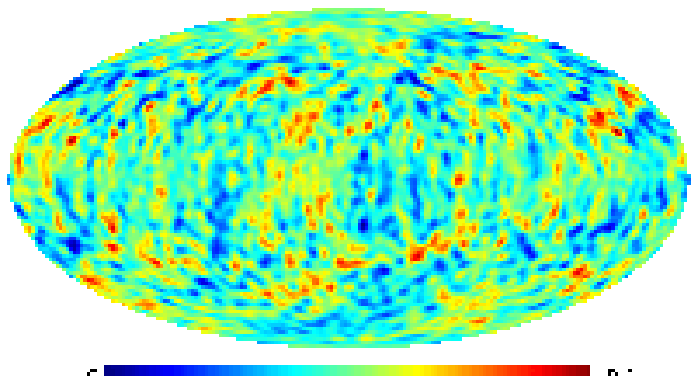}}
\caption{The $D(\theta,\phi)$ maps for PS model. The top panel is the ILC map plus 2 symmetrical PS signal and the bottom is the $D(\theta,\phi)$ map ($\l \le 50$) without Eq.(\ref{eq1b}) generalization. The colorbar limits are $[-0.6,6.3]$mK and $[-0.5,0.5]$mK, respectively.}
\label{ps}
\end{apjemufigure}
An important question is that is the symmetry of the Galaxy image in $\phi$ direction important for extraction of the brightest part of the signal, or
is the effect simply determined by the symmetry of the Galaxy image in $\theta$ direction?
To answer this question, in Fig.\ref{ps5} we plot the result from the $d^{\Delta}_{\l,m}$ estimation in the model with 5 spots located at $\theta_j=\pi/2$ with different amplitudes and different $\phi_j$. As one can see no symmetry in $\phi$ direction was assumed. The result of reconstruction clearly shows that location of the sources in the Galactic plane in $\phi$ direction is crucial. Unlike the model with symmetric location of the spots in Fig.\ref{ps}, now the residuals of the extraction of the spots dominate over the rest of the signal in the Galactic plane. However, as is seen from Fig.\ref{ps5}, the $4n$-correlation of image exists.
\begin{apjemufigure}
\centerline{\includegraphics[width=0.62\linewidth]{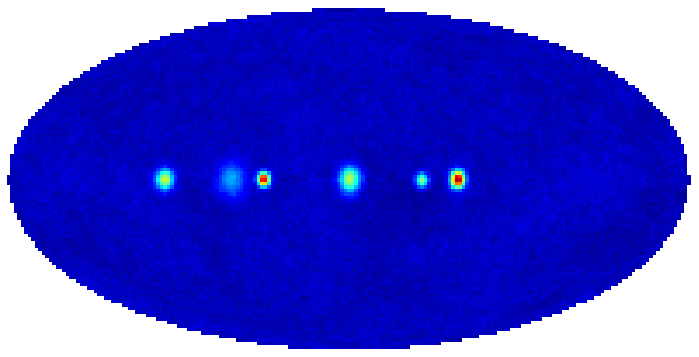}}
\centerline{\includegraphics[width=0.62\linewidth]{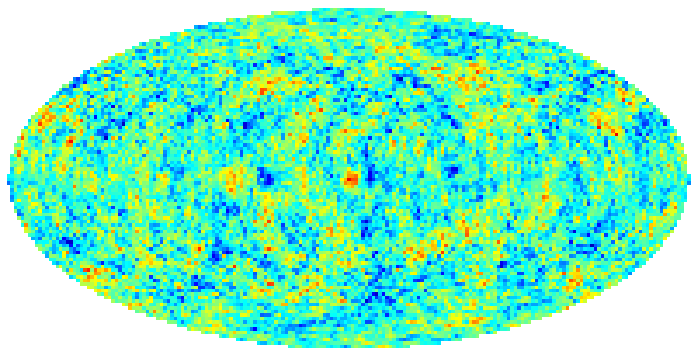}}
\caption{The $D(\theta,\phi)$ maps for another PS model. The top is the ILC map plus 6 PS signal and the bottom is the $D(\theta,\phi)$ maps from the top panel without Eq.(\ref{eq1b}) generalization. The colorbar limits are $[-0.6,6.3]$mK and $[-0.5,0.5]$mK, respectively}
\label{ps5}
\end{apjemufigure}

The properties of the $a_{\l,m}$ coefficients in the spots model are related with the sum (see Eq.(\ref{ps1}))
\begin{equation}
S(m)=\sum_j A_j e^{-im\phi_j}.
\label{aa}
\end{equation}
Actually, Eq.(\ref{aa}) determines the phases of the $a_{\l,m}$ coefficients. Let us discuss the model of two symmetrically situated PS with the same amplitudes $A_1=A_2$ and $\phi_1=\pi/2$, $\phi_2=2\pi-\phi_1$. For this particular case ${\rm Im}[ S(m)]=0$, while ${\rm Re} [S(m)]=2A\cos(\pi m/2) $ .

For $m=2k+1, k=0,1,2\ldots$ we get ${\rm Re} [S(m)]={\rm Im} [S(m)]=0$ and the contribution of the strong signal to the map ILC + two PS vanishes. This means that for amplitudes $a_{\l,m}=c_{\l,m}+p_{\l,m}$, where $c_{\l,m}$ and $p_{\l,m}$ correspond to the ILC and PS signals respectively) and phases $\Psi_{\l,m}$ of $a_{\l,m}$ coefficients, we have
\begin{eqnarray}
a_{\l,m=2k+1}& = & c_{\l,m=2k+1}, \nonumber\\
\Psi_{\l,m} & = & \xi_{\l,m=2k+1}, \hspace{0.5cm }k=0,1,2\ldots
\label{ph}
\end{eqnarray}
where $\xi_{\l,m}$ are the ILC phases.
As one can see, this is a particular example, when strong, but symmetric in $\phi$ direction signal do not contribute to the set of $a_{\lm}$ coefficients at least for the defined range of multipoles.

Let us discuss the other opposite model, in which the number of spots in the galactic plane is no fewer than 2, and their $\phi_j$ coordinates are random in some range $\Phi,2\pi-\Phi$. No specific assumptions about the amplitudes are needed. In this model the sum $S(m)$ in Eq.(\ref{aa}) mostly is represented by $m=0$ modes, $S(0)= \sum_j A_j$, while all $m\neq 0$ modes $S(m)\ll S(0)$ because of randomness of the phases $m\phi_j$. This model, actually is close to the rectangular model, in which
the width of rectangular side in the $\phi$ direction now is $\Phi,2\pi-\Phi$.
At the end of this section we would like to demonstrate, how the symmetry of the Galaxy image in $\phi$ direction can determine the properties of the $D(\theta,\phi$ map. For that we rotate the W-band map by $20^\circ$ along the pole axis and produce the same estimation of $d^{\Delta}_{\l,m}$, as is done for the Galactic reference system. The result of estimation is shown in Fig.\ref{rot}. For comparison, in this Figure we plot the difference and sum between $D(\theta,\phi)$ maps before and after rotation. From these Figures, these new symmetry of the W band map after rotation simply increase the amplitude of signal in Galactic plane zone, especially in the the central part of it.

At the end of this section, we summarize the main results of investigation of the given models of the GF signal.

\begin{itemize}
\item  
For highly symmetrical signal, like the belt model, all $a_{\lm}$ coefficients vanish for the multipole numbers $\l=2n, n=1,2\ldots$, but $\l\delta \ll 1$.  Modification of the $d^{\Delta}_{\lm}$  estimator in form of Eq.(\ref{eq1b}) is crucial to prevent any contribution to the $D(\theta,\phi)$  map from the GF.

\item  Less symmetric model, like the rectangular model, requires $4n$ transition for
the multipole numbers for $d^{\Delta}_{\l,m}$ estimator, which appears for the range of multipoles $\l\delta \ll 1$.
If the resolution of the map we are dealing with is low, $\max \delta \le 1$ that $4n$ correlation appears for all $\l \le \lmax$ and the corresponding $a_{\lm}$ coefficients for GF are in order of magnitude $a_{\lm} \sim A(\l\delta)^3$.
For $\l\delta \gg 1$ the $4n$ correlation of phases does not exist at all.

\item  The amplitude of the GF signal, $A$, and its dependency on $\l$ (like $A(\l)=A_0(\l/\lmin)^{-\beta}$, $\lmin$ is the minimal multipole number for which $A(\lmin)$  achieve the maxima, and $\beta$ is the power index) are crucial for establishing of the $4n$ correlation of phases. Taking asymptotic
$\l\delta \ll 1$ into account, and defining the critical multipole number $\l_{cr}\sim \delta^{-1}$, we can estimate the corresponding amplitudes $A(\l_{cr})=A(1/\delta \lmin)^{-\beta}$. If at that range of $\l$ we get $A(\l_{cr})\ll
C^{\frac{1}{2}}(\l)$, where $C(\l)$ is the CMB power spectrum, the $4n$ correlation would be established for all range of multipoles $\l \le \lmax$, even if it vanishes for the GF signal for $\l\delta \gg 1$. Starting from $\l=\l_{cr}$ and for
$\l > \l_{cr}$ the corresponding $a_{\l,m}$ for GF play a small role in amplitude noise, in comparison to the amplitudes of the CMB signal.

\item  The $d^{\Delta}_{\l,m}$  estimator effectively decreases the amplitudes of the point-like sources located in the Galactic plane, if they have nearly the same amplitudes and are symmetrically distributed in $\phi$ direction around Galactic center. Non-symmetrical and different in amplitudes point-like sources after implementation of $d^{\Delta}_{\l,m}$ estimator produce significant residues.
\end{itemize}

\begin{apjemufigure}
\centerline{\includegraphics[width=0.65\linewidth]{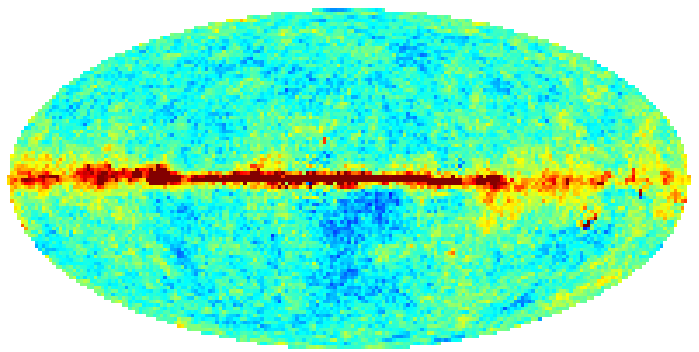}}
\centerline{\includegraphics[width=0.65\linewidth]{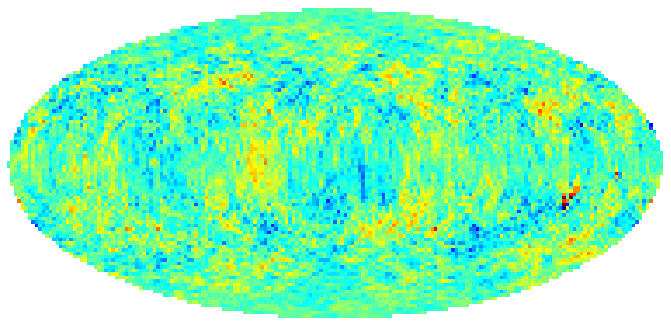}}
\centerline{\includegraphics[width=0.65\linewidth]{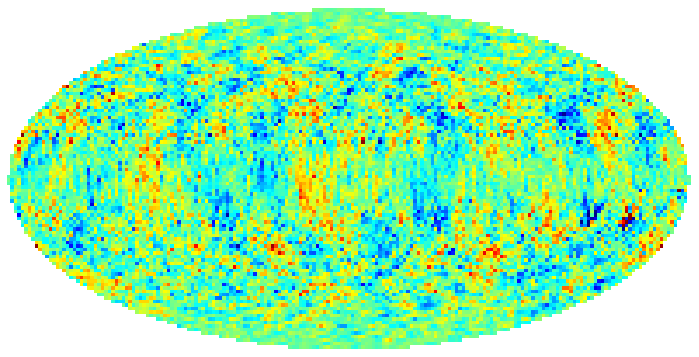}}
\centerline{\includegraphics[width=0.65\linewidth]{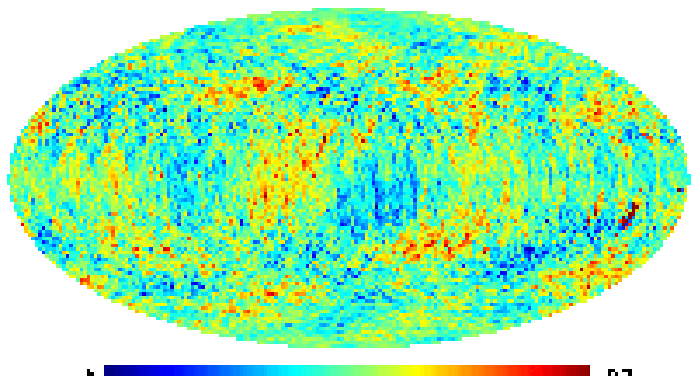}}
\caption{The W band map, rotated by $20^\circ$ in $\phi$ direction (top).
The 2nd is the $D(\theta,\phi)$ map derived from the top panel (without Eq.(\ref{eq1b}) generalization). The 3rd is the difference between $D(\theta,\phi)$ maps from W band before and after rotation. The bottom is the sum between $D(\theta,\phi)$ maps (before and after rotation). The colorbar limits are $[-0.5,0.5]$mK for all panels.}
\label{rot}
\end{apjemufigure}

\section{Symmetry of the \wmap foregrounds}
In this section we apply the proposed $d^{\Delta}_{\l,m}$ estimator to the maps for \wmap Q, V and W band foregrounds (which are sum of synchrotron, free-free and dust emission). We then transform them by the $d^{\Delta}_{\l,m}$ estimator. These foreground maps do not contain the CMB signal and instrumental noise, therefore they allow us to estimate the properties of the GF in details.
In Fig.\ref{fgd} we plot the $D(\theta,\phi)$ maps for Q, V and W band foregrounds ($\Delta=4$) for the multipole range $\l\le 46$. This range is determined by the resolution of the \wmap foregrounds maps ($\l\le 50$). As one can see from these maps, the GF perfectly follows to $4n$ multipole correlation, which remove the brightest part of the signal down to the level $\pm$ 50 mK for the Q band, $-0.19,0.50$ mK for the V band , $-0.09,0.29$ mK for the W band and
$-0.1,0.1$ mK for the $D(\theta,\phi)$ map, the difference between V and W foregrounds. Note that these limits are related with the brightest positive and negative spots (point sources) in the maps, while diffuse components have significantly smaller amplitudes.
To show the high resolution $D(\theta,\phi)$ map which characterizes the properties of the foregrounds in V and W band, in Fig.\ref{pow} we plot the map of difference $V-W$ bands, and the corresponding $D(\theta,\phi)$ map for $\l\le 250$. Note that $V-W $ map does not contain the CMB signal, but for high $\l$ the properties of the signal are determined by the instrumental noise.

\section{The power spectrum and correlations of the $D$ map} \label{power}
\subsection{What is constructed from the $d^{\Delta}_{\l,m}$ estimator?}

To characterize the power spectrum of the $D(\theta,\phi)$ maps we introduce the definition
\begin{eqnarray}
D(\l)=\frac{1}{2\l+1}\sum_{m=-\l}^\l |d^{\Delta}_{\l,m}|^2.
\label{pp}
\end{eqnarray}
If the derived $d^{\Delta}_{\l,m}$ signal is Gaussian, that power represents all the statistical properties of the
signal. For non-Gaussian signal, $D(\l)$ power characterizes the diagonal elements of the correlation matrix.
From Fig.\ref{Psim} it can be clearly seen that for \wmap foregrounds, especially for V and W bands, the power spectra of
$D(\theta,\phi)$ are significantly smaller than the power of the CMB, for estimation of which we simply use the power of TOH FCM map, transformed by $d^{\Delta}_{\l,m}$ estimator as
\begin{equation}
D_{\rm fcm}(\l)=\frac{1}{2\l+1}\sum_{m=-\l}^\l |c_{\l,m}-\frac{|c_{\l,m}|}{|c_{\l+\Delta,m}|}c_{\l+\Delta,m}|^2,
\label{pp1}
\end{equation}
assuming that FCM map is fairly clean from the foreground signal. An important point of analysis of the
\wmap foregrounds is that for V and W bands $d^{\Delta}_{\l,m}$ estimator decreases significantly the amplitude of GF, practically by 1 to 2 order of magnitude below the CMB level.

The most intriguing question related to $4n$-correlation of the derived map from the \wmap V and W band signals is what is reproduced by the $d^{\Delta}_{\l,m}$ estimator?
The next question, which we would like to discuss is why the power spectrum of  $d^{\Delta}_{\l,m}$ estimation of the V and W bands shown in Fig.\ref{fig22} are practically the same at the range of multipoles $\l \le 100$, when we can neglect the contribution from instrumental noise to both channels and differences of the antenna beams. The equivalence of the powers for these two signals, shown in Fig.\ref{fig22}, clearly tell us that these derived maps are related with pure CMB signal (which we assume to be frequency independent).

\begin{apjemufigure}
\centerline{\includegraphics[width=0.65\linewidth]{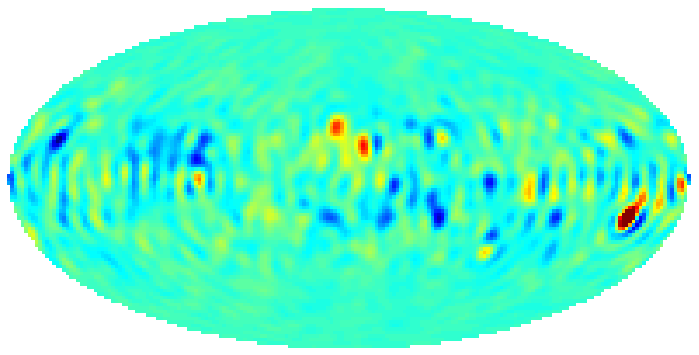}}
\centerline{\includegraphics[width=0.65\linewidth]{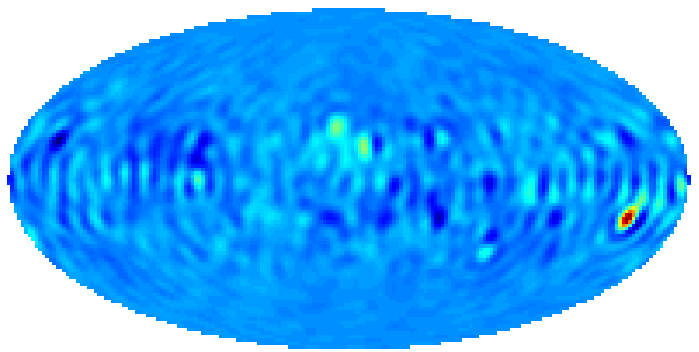}}
\centerline{\includegraphics[width=0.65\linewidth]{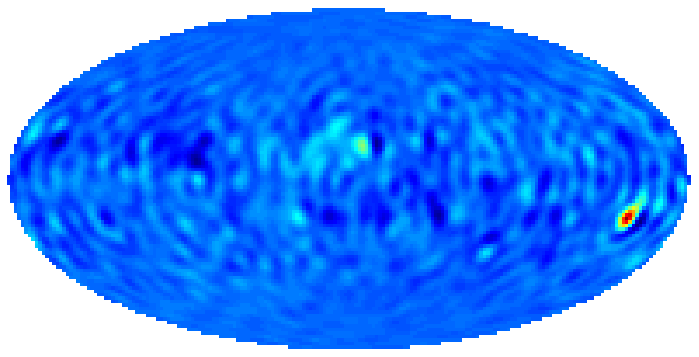}}
\centerline{\includegraphics[width=0.65\linewidth]{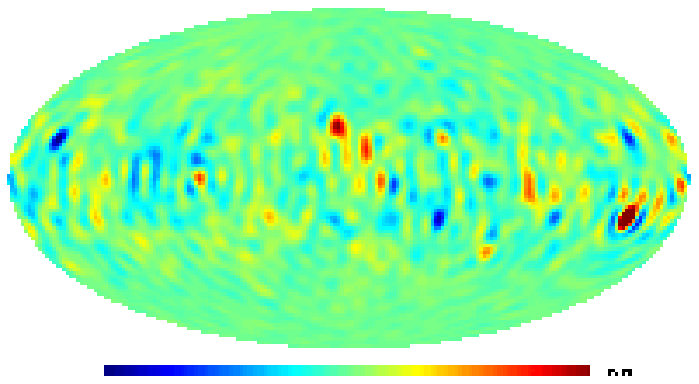}}
\caption{(From top to bottom the $D(\theta,\phi)$ map for Q, V and W band foregrounds, respectively, with $\lmax=46$, $\Delta=4$. The bottom is the $D(\theta,\phi)$ map from the V and W band map difference: $V-W$.}
\label{fgd}
\end{apjemufigure}
\begin{apjemufigure}
\centerline{\includegraphics[width=0.65\linewidth]{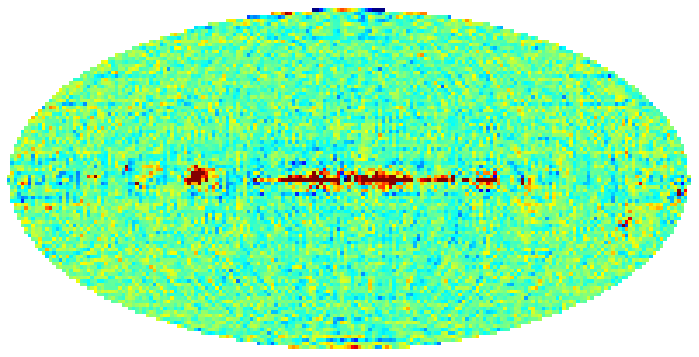}}
\centerline{\includegraphics[width=0.65\linewidth]{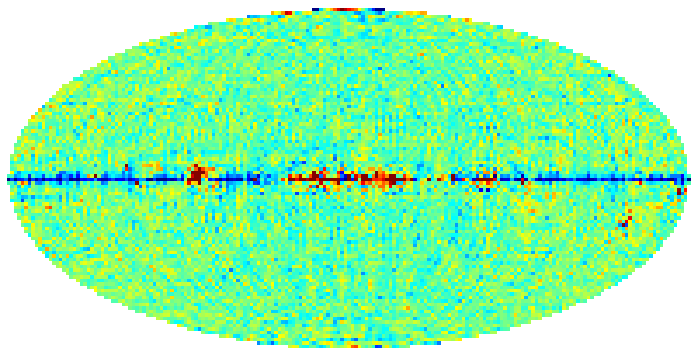}}
\centerline{\includegraphics[width=0.65\linewidth]{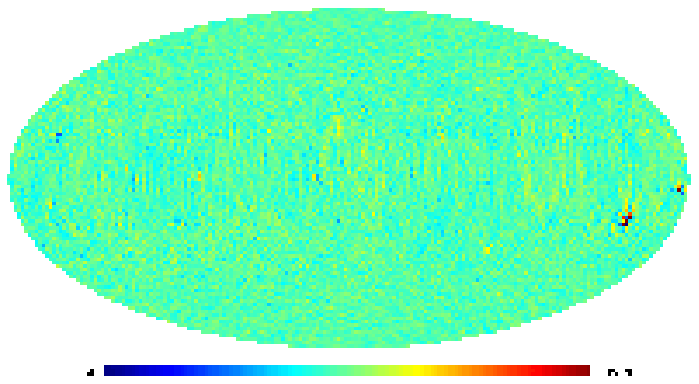}}
\caption{The 1st panel is the difference between V and W bands: V$-$W, the 2nd is the same map, but with all $m=0$ modes in $\alm$ being set zero, and the 3rd is the $D(\theta,\phi)$ map from the 1st panel.}
\label{pow}
\end{apjemufigure}
\begin{apjemufigure}
\centerline{\includegraphics[width=0.65\linewidth]{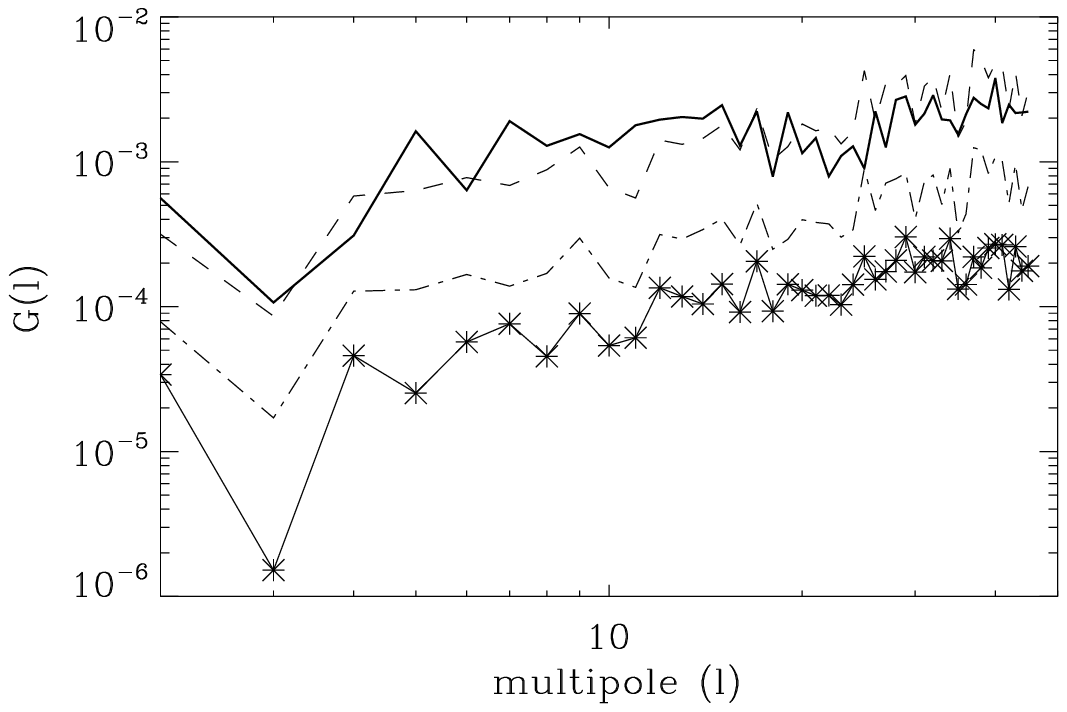}}
\caption{The power spectrum $G(\l)=D(\l) \l(\l+1)/2\pi$ for $D(\theta,\phi)$ map (W band, the solid line with stars) in comparison with the $G(\l)$ power of the $D(\theta,\phi)$ map for FCM map (the thick solid line). The dash line represents the power spectrum $G(\l)$ for the $D(\theta,\phi)$ map for the difference between Q and W bands: Q$-$W, and the dash-dot line is for V$-$W.
}
\label{Psim}
\end{apjemufigure}

In this section we present some analytical calculations which clearly demonstrate what kind of combinations between amplitudes and phases of the CMB signal in the
V, W bands and phases of foregrounds are represented in the $d^{\Delta}_{\l,m}$ estimator. As  was mentioned in
Section 1, this estimator is designed as a linear estimator of the phase difference $\Phi_{\l+\Delta,m}-\Phi_{\l,m}$, if the phase difference is small. Let us introduce the model of the signal at each band $a^{(j)}_{\l,m}=c_{\l,m}+F^{(j)}_{\l,m}$, where $c_{\l,m}$ is frequency independent CMB signal and $F^{(j)}_{\l,m}$ is the sum over all kinds of foregrounds for each band $j$ (synchrotron, free-free, dust emission etc.).

\begin{apjemufigure}
\centerline{\includegraphics[width=0.65\linewidth]{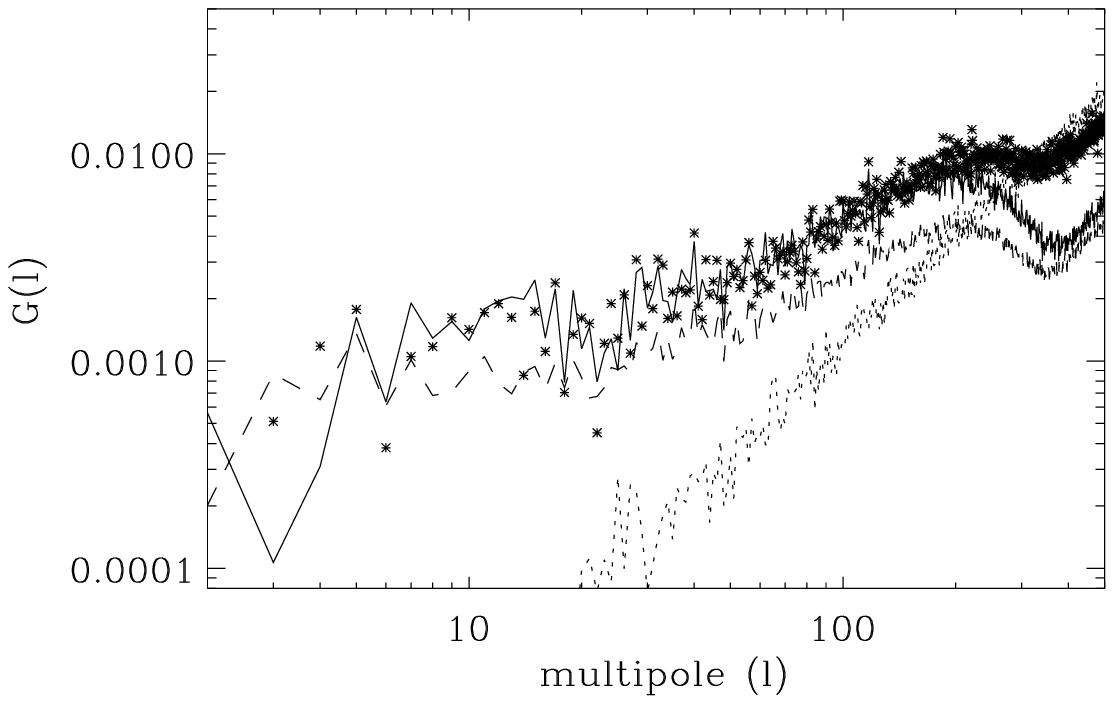}}
\centerline{\includegraphics[width=0.65\linewidth]{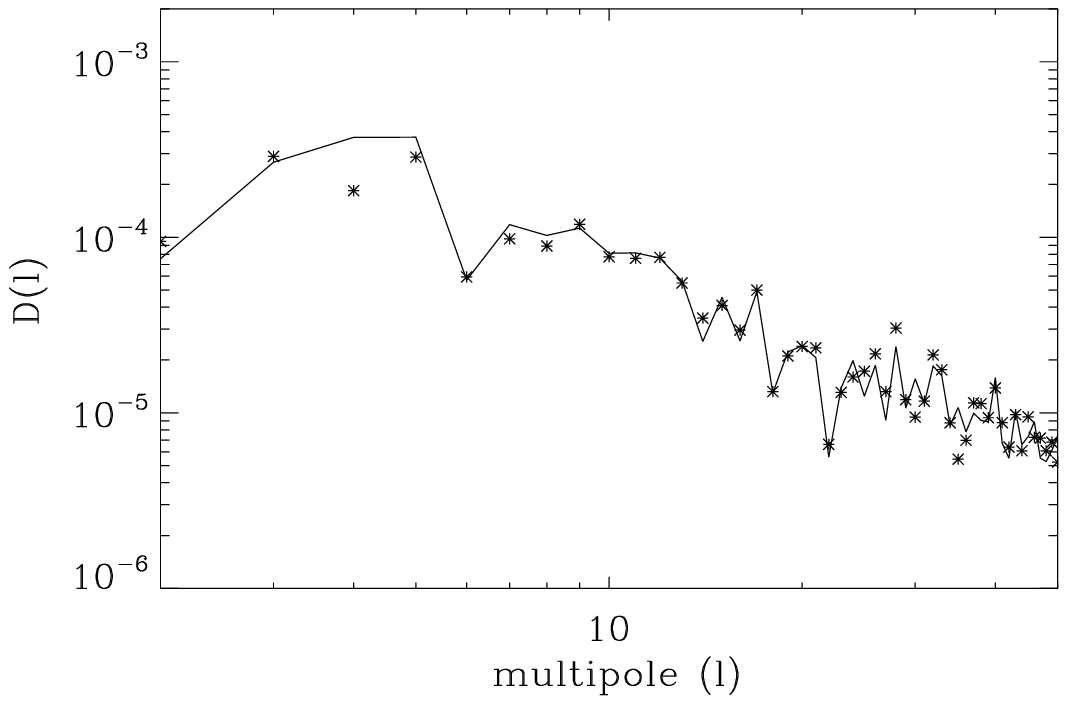}}
\caption{The power spectrum $G(\l)=D(\l) \l(\l+1)/2\pi$ for $D$ map (W band, stars) in comparison to the powers of the ILC map (dash) and the FCM (solid line). The dotted line represents the $G$ power of the D map for difference V$-$W. The bottom is the low resolution power spectra $D(\l)$ for W and V bands.}
\label{fig22}
\end{apjemufigure}

According to the investigation above on the foreground models, it is realized that without the ILC signal the $d^{\Delta}_{\l,m}$ estimation of the foregrounds, especially for V and W bands, corresponds to the signal \footnote{Hereafter we omit the mark of channel $j$ to simplify the formulas}
\begin{equation}
d^{\Delta,(f)}_{\l,m}=F_{\l,m}-\frac{|F_{\l,m}|}{|F_{\l+\Delta,m}|}F_{\l+\Delta,m},
\label{ff}
\end{equation}
the power of which is significantly smaller then that of the CMB
\begin{equation}
d^{\Delta,(cmb)}_{\l,m}=c_{\l,m}-\frac{|c_{\l,m}|}{|c_{\l+\Delta,m}|}c_{\l+\Delta,m}.
\label{cmb}
\end{equation}
In terms of moduli and phases of the foregrounds at each frequency band

\begin{eqnarray}
F_{\l,m}&=&|F_{\l,m}|\exp(i\Phi_{\l,m}), \nonumber\\
c_{\l,m}&=&|c_{\l,m}|\exp(i\xi_{\l,m}),
\label{dd0}
\end{eqnarray}
where $\Phi_{\l,m}$ and $\xi_{\l,m}$ are the phases of foreground and the CMB, respectively. And from Eq.(\ref{dd0}) we get
\begin{equation}
d^{\Delta,(f)}_{\l,m}=|F_{\l,m}|\left(e^{i\Phi_{\l,m}}-e^{i\Phi_{\l+\Delta,m}}\right),
\label{ff}
\end{equation}
and practically speaking, we have $\Phi_{\l,m}=\Phi_{\l+\Delta,m}$ .
Thus, taking the $4n$ correlation into account, we can conclude that it reflects directly the high correlation of the phases of the foregrounds, determined by the GF. Moreover, if any foreground cleaned CMB maps derived from different methods display the $4n$ correlation of phases, it would be evident that foreground residuals still determine the statistical properties of the derived signal.

\subsection{$4n$ phase correlation of the $D$ map}
One of the basic ideas for comparison of phases of two signals is to define the following trigonometric moments
for the phases $\xi_{\l^{'},m}$  and $\Psi_{\l,m}$ as:
\begin{eqnarray}
\Cs(\l,\l^{'})=\frac{1}{\sqrt{\l}}\sum_{m=1}^{\l} \cos\left(\xi_{\l^{'},m}-\Psi_{\l,m}\right);\nonumber\\
\Si(\l,\l{'})=\frac{1}{\sqrt{\l}}\sum_{m=1}^{\l} \sin\left(\xi_{\l^{'},m}-\Psi_{\l,m}\right),\nonumber\\
\label{def2}
\end{eqnarray}
where $\l \le \l^{'}$. We apply these trigonometric moments to investigate the phase correlations for TOH FCM and WFM. For that we simply substitute $\l= \l^{'}$  in Eq.(\ref{def2}), and define $\xi_{\l,m}$ as the phase of FCM and $\Psi_{\l,m}$ as that of WFM. The result of the calculations is presented in Fig.\ref{comp}.

From Fig.\ref{comp} it can be clearly seen that the FCM has strong $\Delta \l=4$ correlations starting from $\l\simeq 40$ which rapidly increase for $\l > 40$,  while for WFM these correlations are significantly damped, especially at low multipole range $\l \le 40$.
However, the $d^{\Delta}_{\l,m}$ estimator allow us to clarify the properties of phase correlations for low multipole range. The idea is to apply $d^{\Delta}_{\l,m}$ estimator to FCM and WFM, and to compare the power spectra of the signals obtained before and after that.
According to the definition of $d^{\Delta}_{\l,m}$ estimator, the power spectrum of the signal is given by Eq.(\ref{pp1}), which now has the form
\begin{equation}
D(\l)=\frac{2}{\l}\sum_m|c_{\l,m}|^2\left[1-\cos(\xi_{\l+\Delta,m}-\xi_{\l,m})\right].
\label{dd}
\end{equation}

\begin{apjemufigure}
\centerline{\includegraphics[width=0.85\linewidth]{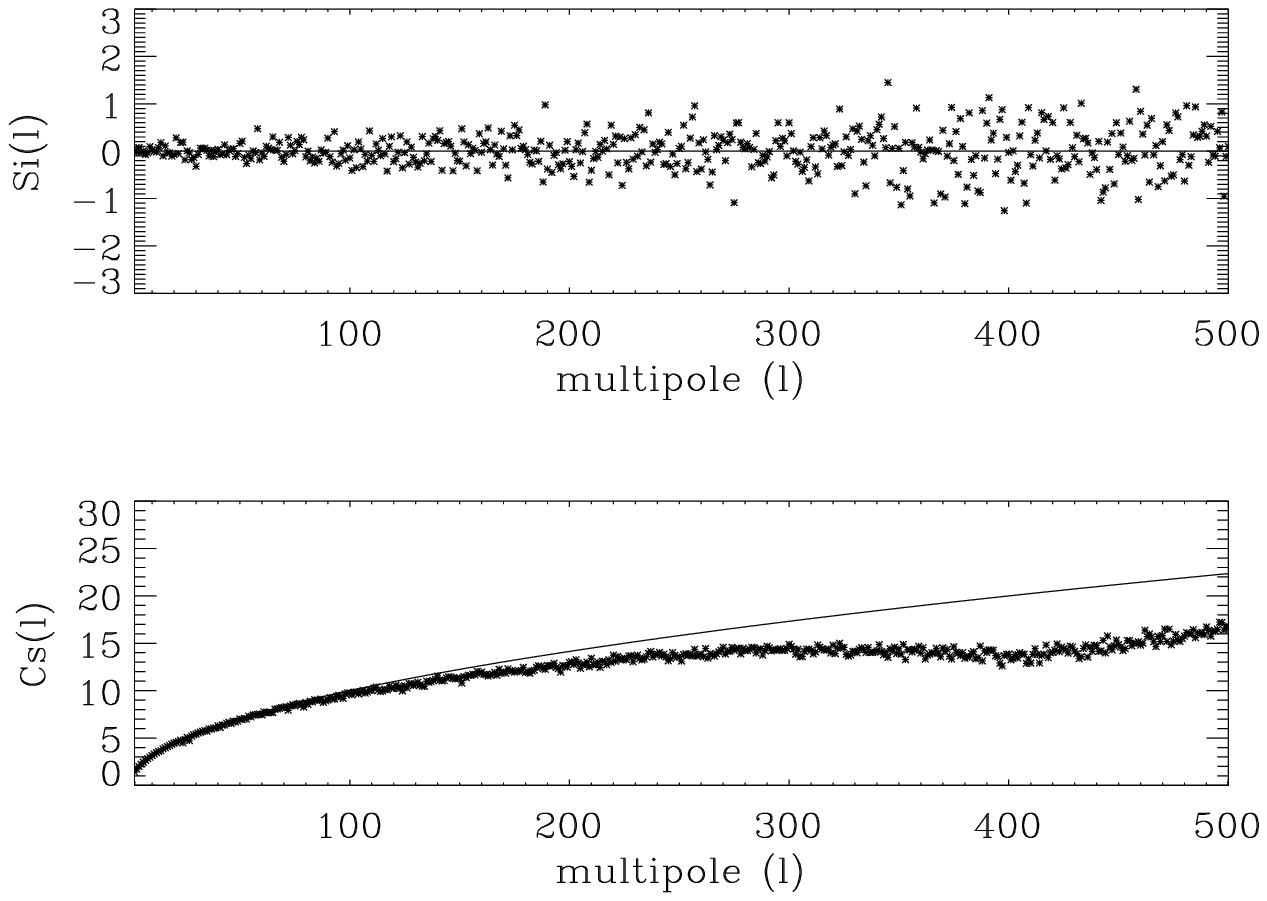}}
\centerline{\includegraphics[width=0.85\linewidth]{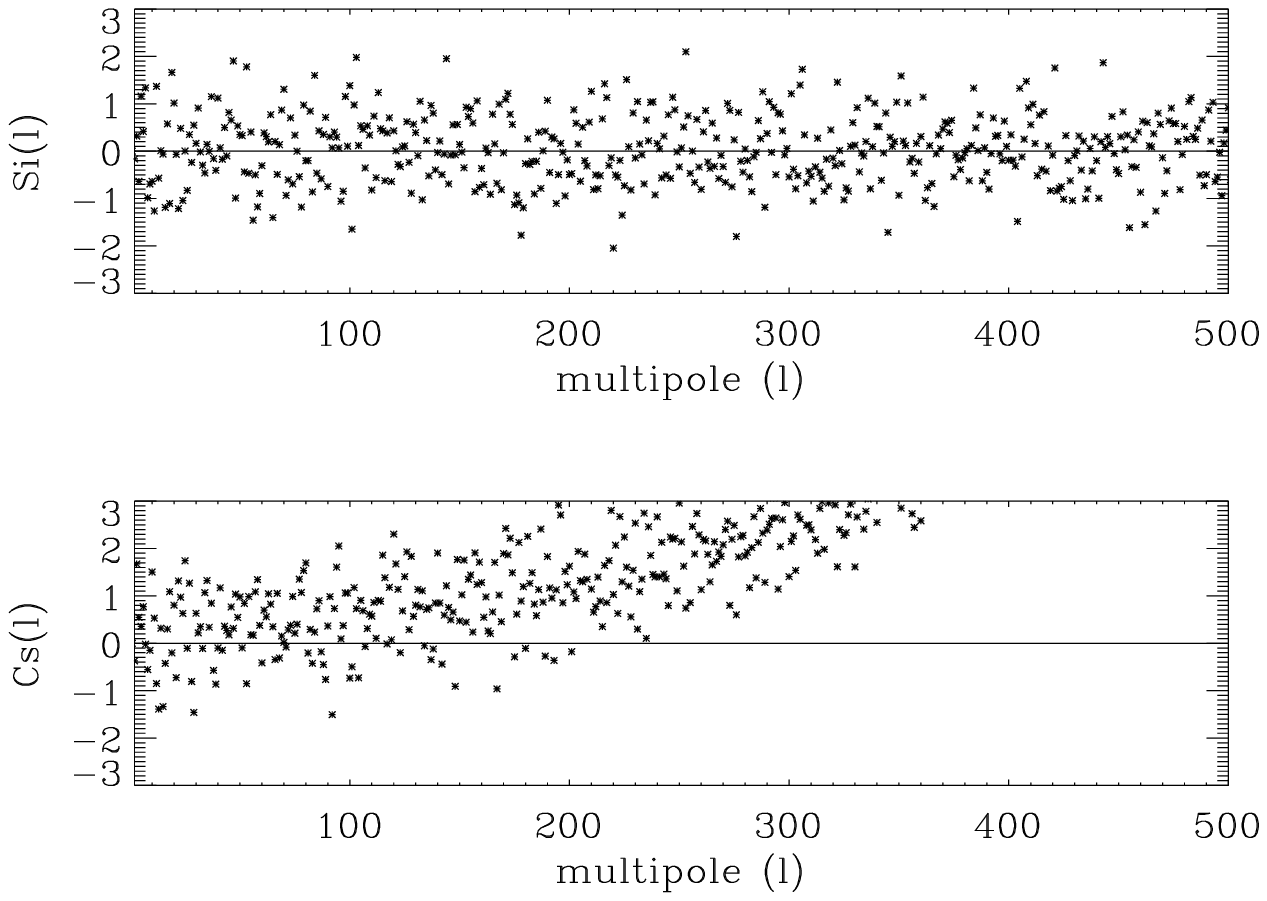}}
\centerline{\includegraphics[width=0.85\linewidth]{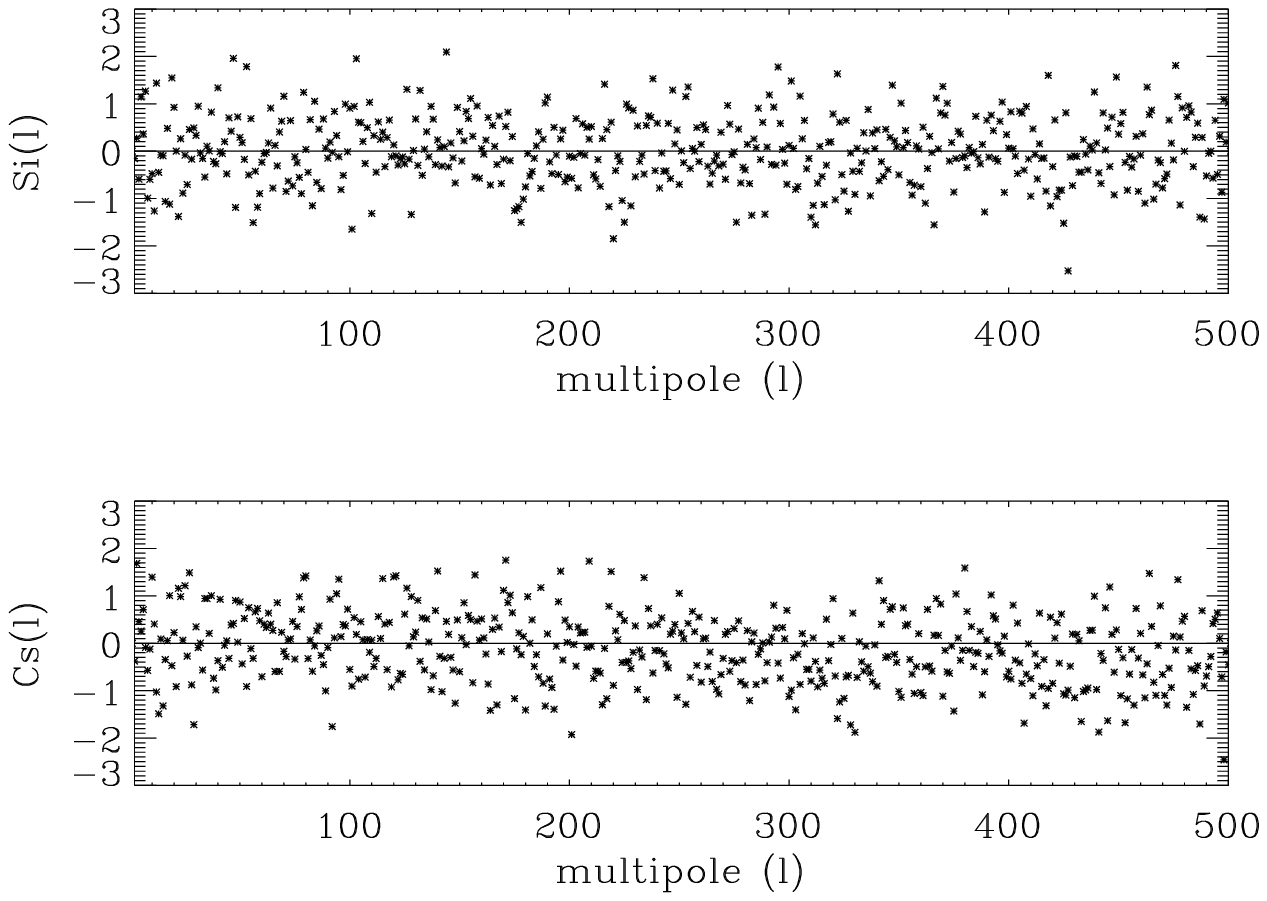}}
\caption{The $\Cs(\l)$ and $\Si(\l)$ trigonometric moments for the cross-correlation of phases between the TOH FCM and WFM (the first pair). The solid line represents the limit when the phases are identical. The middle pair is The $\Cs(\l)$ and $\Si(\l)$ trigonometric moments for the FCM with phase difference $\xi_{\l+4,m}-\Psi_{\l,m}$. The bottom pair is 
$\Cs(\l)$ and $\Si(\l)$ trigonometric moments for the WFM with phase difference $\xi_{\l+4,m}-\xi_{\l,m}$.}
\label{comp}
\end{apjemufigure}

The last term in Eq.(\ref{dd}) corresponds to the cross-correlation between $\l,m$ and $\l+4,m$ modes, which should vanish for Gaussian random signals after averaging over the realization. For a single realization of the random Gaussian process this term is non-zero because of the same reason, as well known ``cosmic variance'', implemented for estimation of the errors of the power spectrum estimation (see Naselsky et al. 2004). Thus
\begin{equation}
D(\l)\simeq\frac{2}{\l}\sum_m|c_{\l,m}|^2,
\label{dd1}
\end{equation}
and error of $D(\l)$ is in order to
\begin{equation}
\frac{\Delta D(\l)}{D(\l)}\simeq\frac{2}{\sqrt{\l+\frac{1}{2}}}.
\label{dderror}
\end{equation}
To evaluate qualitatively the range of possible non-Gaussianity of the FCM and WFM, in Fig.\ref{Fpow} we plot the function $F(\l)=2 [D(\l)-2C(\l)]/[D(\l)+C(\l)]$ for FCM and WFM, in which we mark the limits $\pm 2/\sqrt{\l}$.
As one can see, potentially dangerous range of low multipoles is $\l=3,4$, $\l=21-24$, $\l \simeq 100-150$ for the WFM. Non-randomness on some of the multipole modes is mentioned in \citet{autocross}.

\begin{apjemufigure}
\centerline{\includegraphics[width=0.65\linewidth]{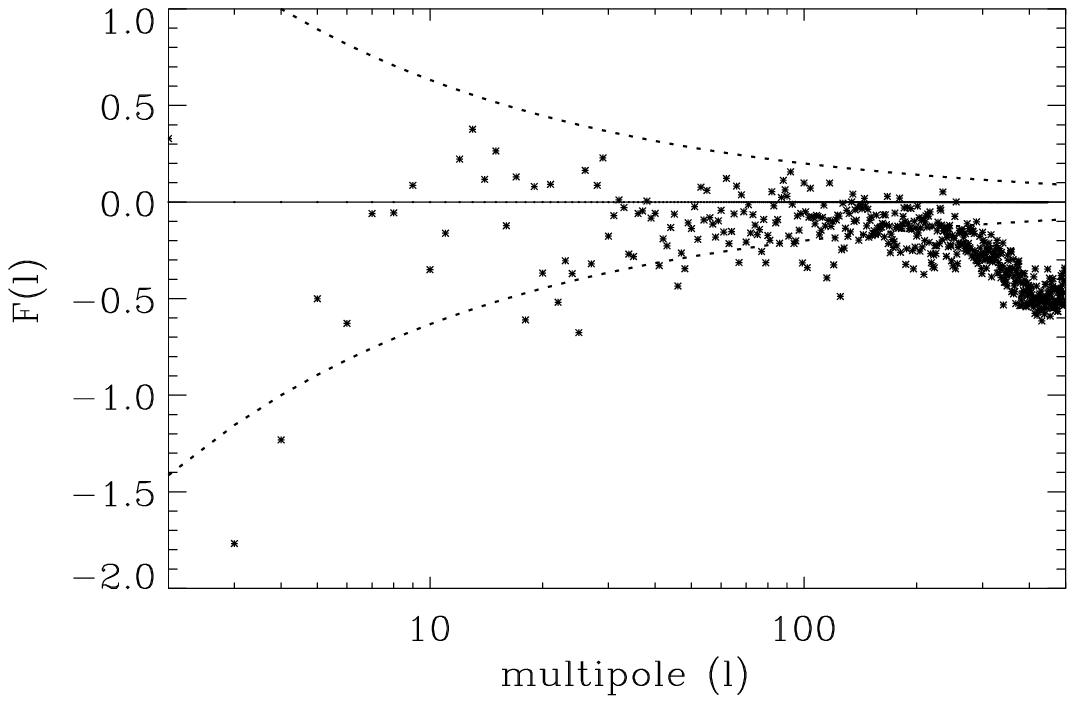}}
\centerline{\includegraphics[width=0.65\linewidth]{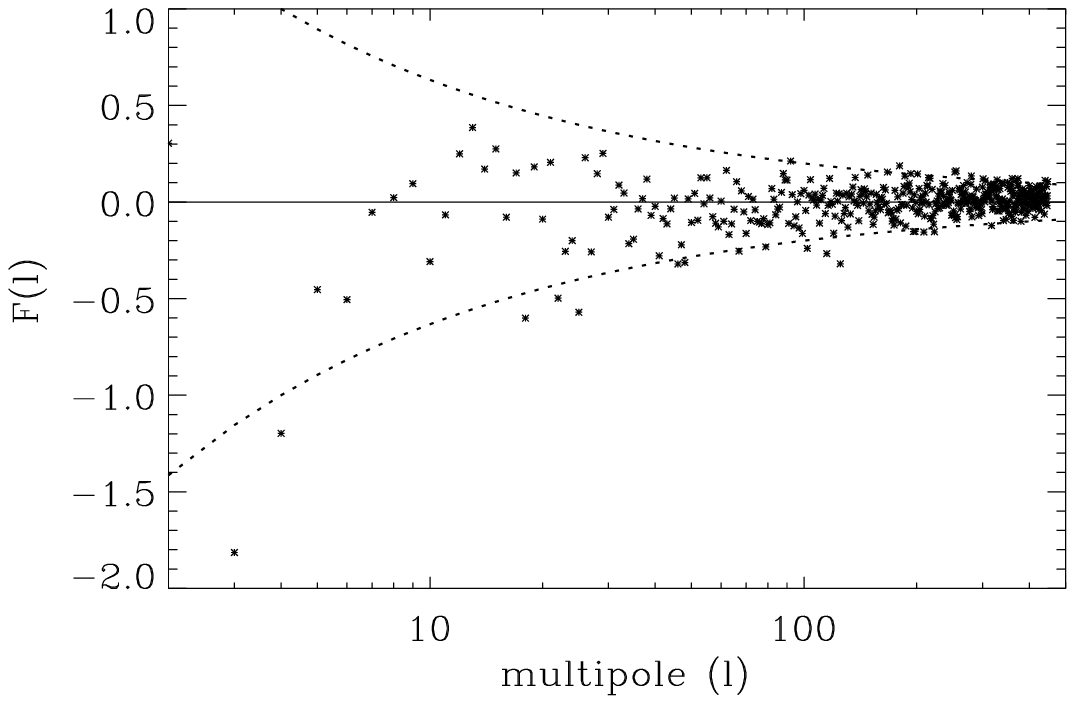}}
\caption{$F(\l)$ function for the TOH FCM (top) and the WFM (bottom). The dotted lines represent $\pm 2/\sqrt{\l}$ limit.}
\label{Fpow}
\end{apjemufigure}

At the end of this section we would like to demonstrate that application of $d^{\Delta}_{\l,m}$ estimator to maps with foregrounds residuals, such as the FCM, provides additional ``cleaning''. In Fig.\ref{clean} we present the $\Cs(\l)$ and $\Si(\l)$ trigonometric moments for the FCM with shift of the multipoles $\l^{'}=\l +2$. One can see that the $\Delta =2$ correlation of phases is strong (practically, they are at the same level as $\Delta =4$ correlations). However, after $d^{\Delta}_{\l,m}$ filtration these correlations are significantly decreased.

The implementation of the $d^{\Delta=4}_{\l,m}$ estimator to the non-Gaussian signal significantly decreases these correlations.

The properties of the  $d^{\Delta}_{\l,m}$ estimator described can manifest themselves more clearly in terms of images of the CMB signal. In Fig.\ref{clean1} we plot the results of the maps with $d^{\Delta}_{\l,m}$ implemented on FCM and WFM, in order to demonstrate how the estimator works on the non-Gaussian tails of the derived CMB maps. In Fig.\ref{clean1} we can clearly see that the morphology of the $D(\theta,\phi)$ maps are the same
and difference between $D_{\rm fcm}(\theta,\phi)$ and $D_{\rm wfm}(\theta,\phi)$ is related to point sources residuals localizes outside the galactic plane (see the 3rd panel). A direct substraction of the WFM from the FCM reveals significant contamination of the GF residuals and non -galactic point sources ( the third from the bottom and bottom maps). The second from the bottom map corresponds to difference between
$D_{\rm fcm}(\theta,\phi)$ and $D_{\rm wfm}(\theta,\phi)$ for which the amplitudes of the signal represented in colorbar limit $\pm 0.1$ mK. One can see that the GF is removed down to the noise level. In combination of the phase analysis we can conclude that the implementation of the $d^{\Delta}_{\l,m}$ estimator looks promising as an additional cleaning  of the GF residuals and can help investigate the statistical properties of derived CMB signals in more detailed.

\begin{apjemufigure}
\centerline{\includegraphics[width=0.85\linewidth]{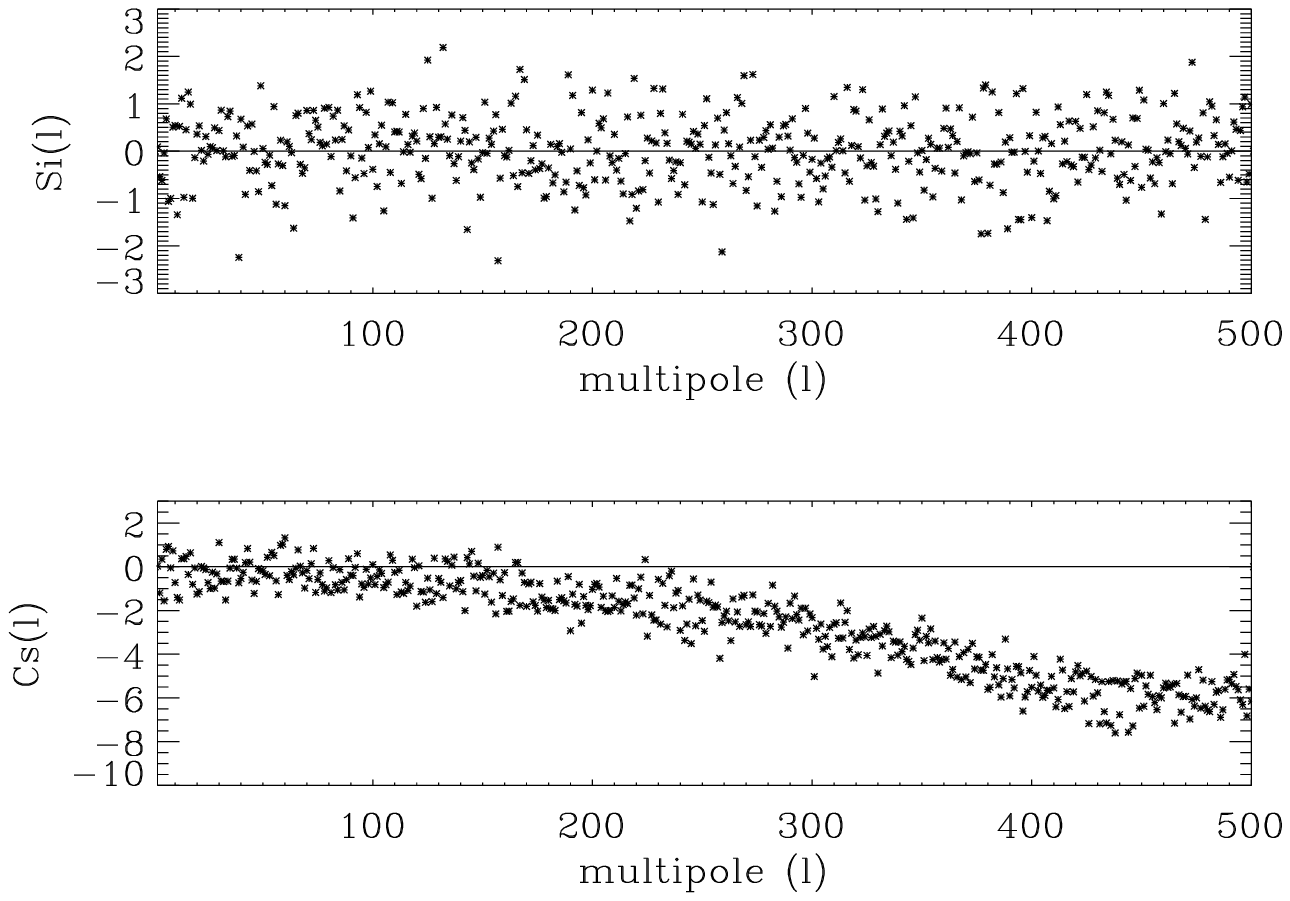}}
\centerline{\includegraphics[width=0.85\linewidth]{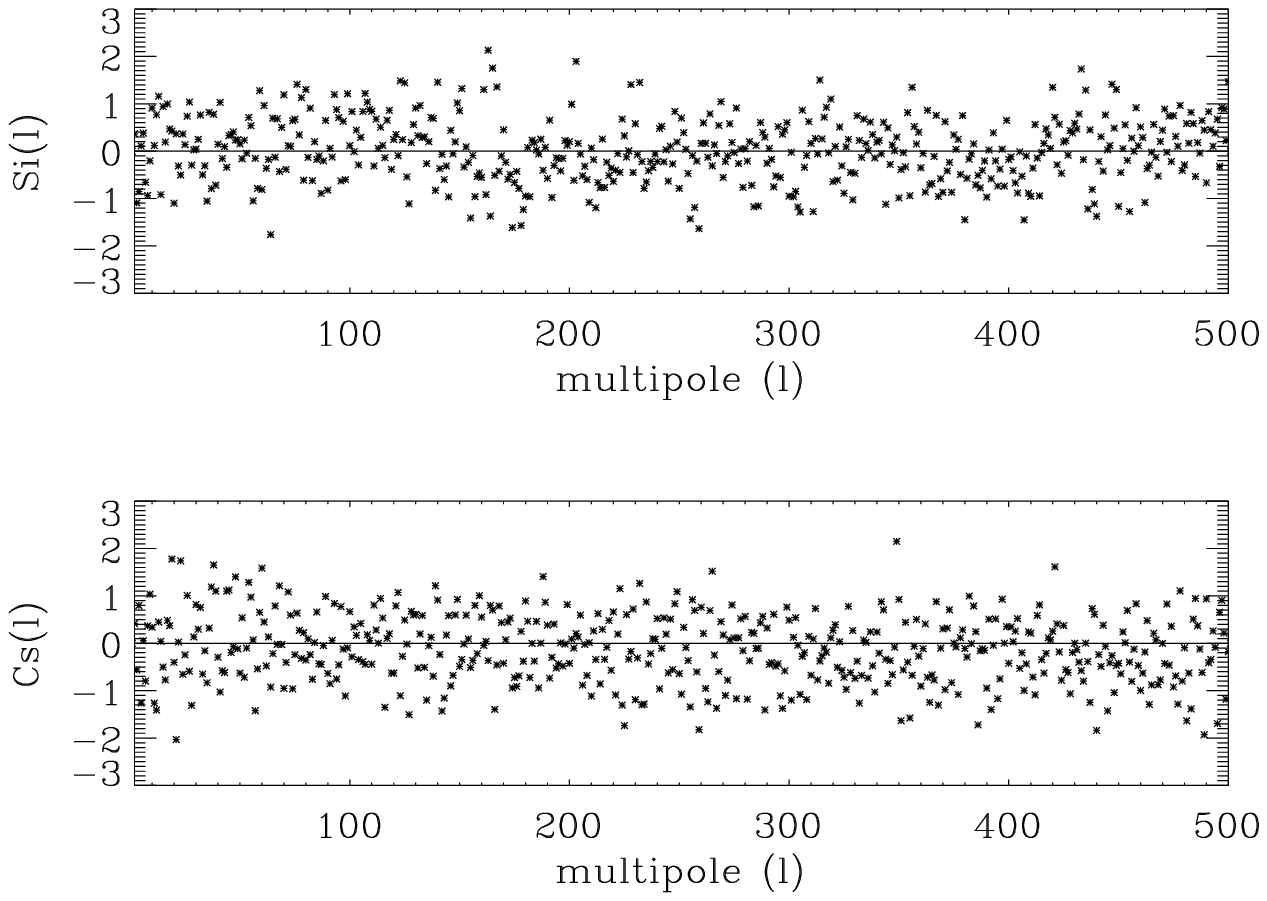}}
\centerline{\includegraphics[width=0.85\linewidth]{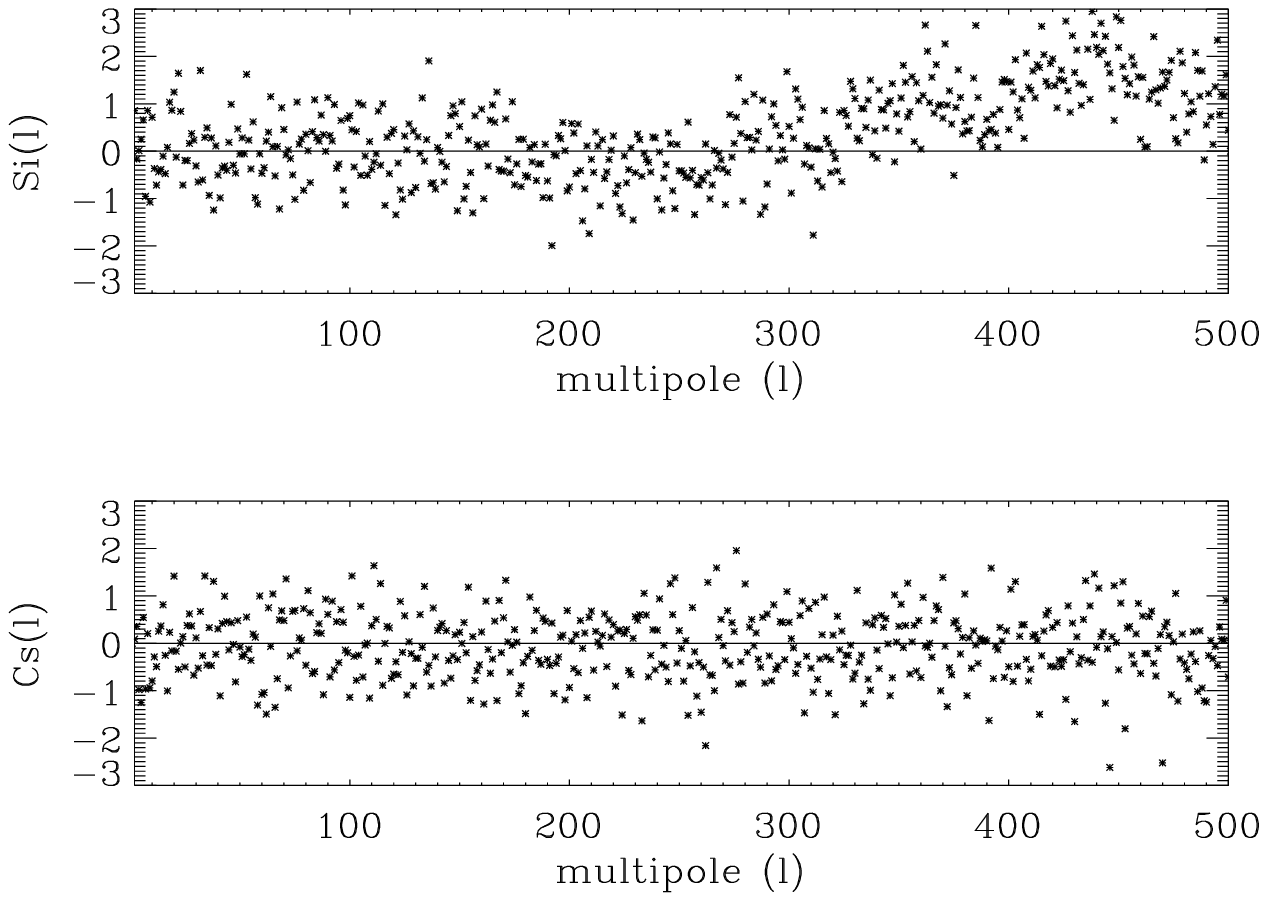}}
\centerline{\includegraphics[width=0.85\linewidth]{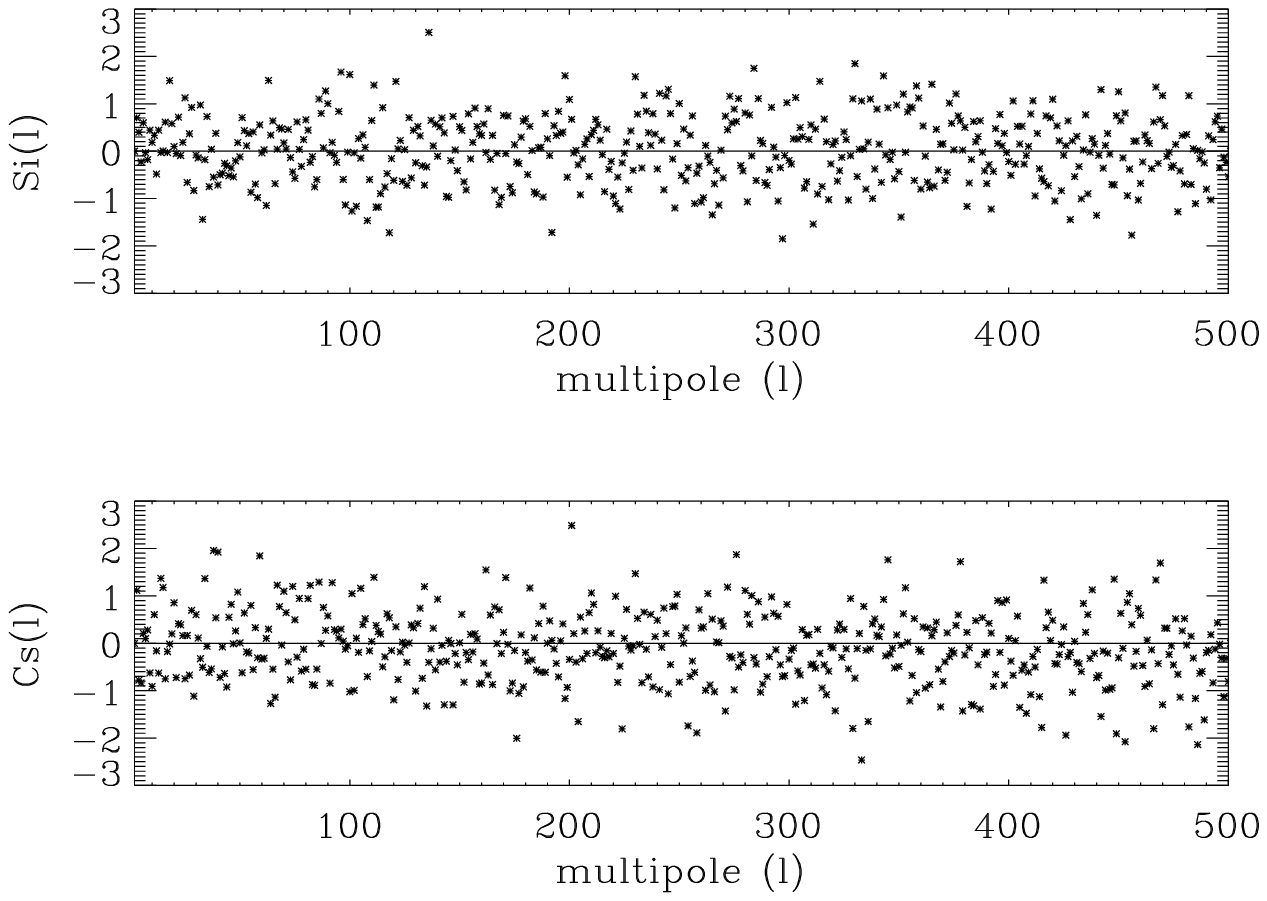}}
\caption{The $\Cs(\l)$ and $\Si(\l)$ trigonometric moments for the FCM at $\Delta \l=2$ (the first pair). The 2nd pair is for $\Delta \l=2$ after $d^{\Delta=4}_{\l,m}$ filtration.
The 3rd and 4th pairs are for $\Delta \l=1$ before and after $d^{\Delta=4}_{\l,m}$ filtration, respectively.}
\label{clean}
\end{apjemufigure}

\begin{apjemufigure}
\centerline{\includegraphics[width=0.65\linewidth]{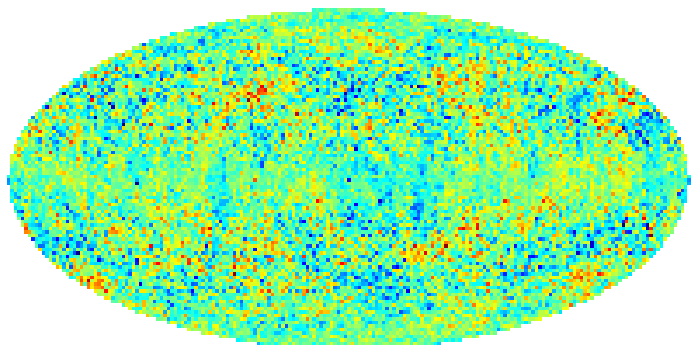}}
\centerline{\includegraphics[width=0.65\linewidth]{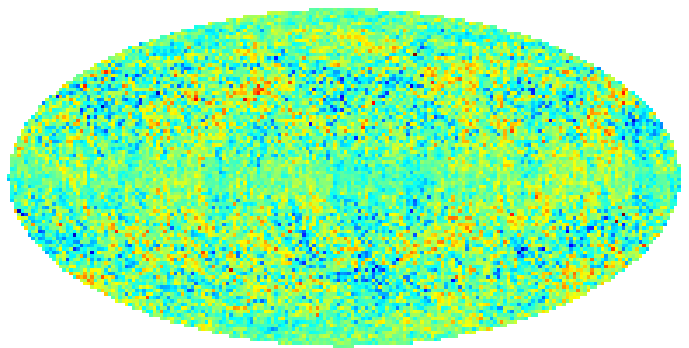}}
\centerline{\includegraphics[width=0.65\linewidth]{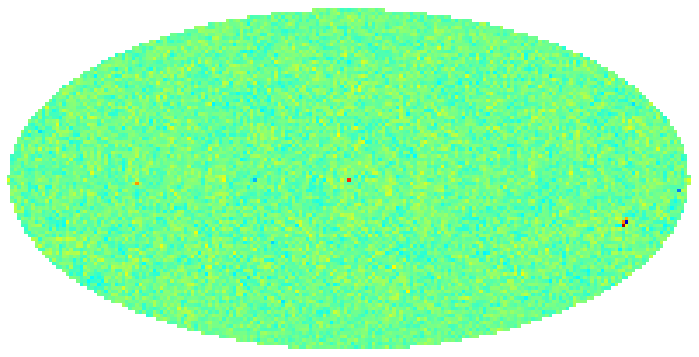}}
\centerline{\includegraphics[width=0.65\linewidth]{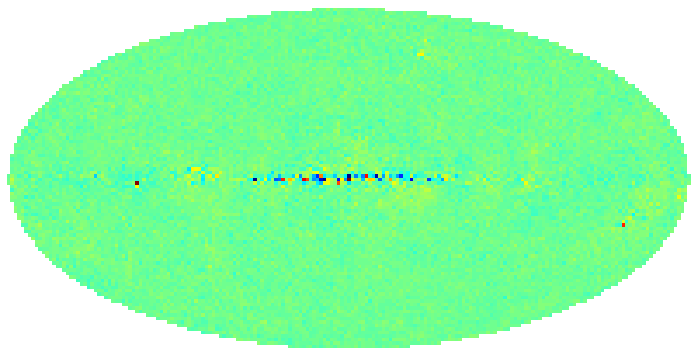}}
\centerline{\includegraphics[width=0.65\linewidth]{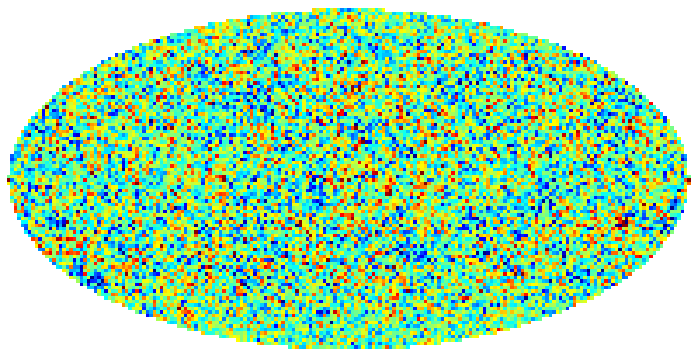}}
\centerline{\includegraphics[width=0.65\linewidth]{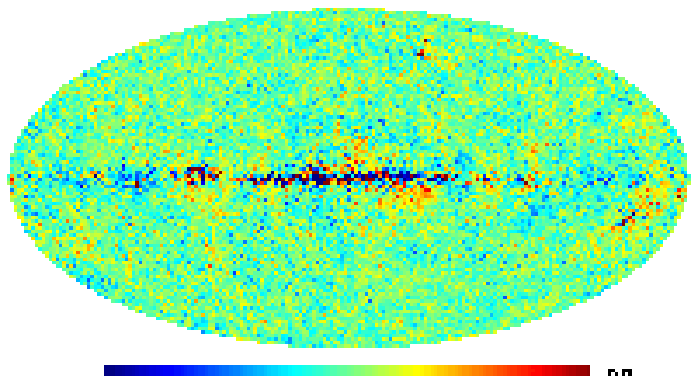}}

\caption{The comparison between the FCM and WFM before and after implementation of the $d^{\Delta}_{\l,m}$ estimator. From the top to the bottom: $D_{\rm fcm}(\theta,\phi)$ map for the FCM, $D_{\rm wfm}(\theta,\phi)$ for the WFM, the difference between $D_{\rm fcm}(\theta,\phi)$ and $D_{\rm wfm}(\theta,\phi)$, and the difference between FCM and WFM. All these maps are plotted with colorbar limit $\pm 0.5$ mK. The last pair are the same as previous pair, but for the colorbar limit $\pm 0.1$ mK. For all the maps $\lmax=500$.}
\label{clean1}
\end{apjemufigure}

\section{Conclusion}
In this paper we examine a specific group of correlations between $\l$, which is used as an estimation of the statistical properties of the foregrounds in the \wmap maps. These correlations, in particular, among phases are closely related to symmetry of the GF (in Galactic coordinate system). An important point of analysis is that for the foregrounds the correlations of phases for the total foregrounds at V and W bands have specific shape when
$\Phi_{\l,m} \simeq \Phi_{\l+\Delta,m}, \Delta =4n, n=1,2,3\ldots$. These correlations can be clearly seen in the W band of the \wmap data sets down to $\lmax=512$ and must be taken into account for modeling of the foreground properties for the upcoming \planck mission.
We apply the $d^{\Delta}_{\l,m}$ estimator to the TOH FCM, which contains strong residuals  from the GF and show that these residuals are removed from the $D_{\rm wfm}(\theta,\phi)$ map.
Moreover, in that map the statistics of the phases display the Gaussian statistics closer than the original FCM (no correlation of phases between different ${\l,m}$ modes except between $\l+\Delta,m$ and $\l,m$, which is chosen as a basic one, defined by the form of $d^{\Delta}_{\l,m}$ estimator.)

In this paper we do not describe in details the properties of the signal derived by $d^{\Delta}_{\l,m}$ estimator from the \wmap V and W bands. Further developments of the method, including multi-frequency combination of the maps and CMB extraction by the estimator will be in a separate paper. To avoid misunderstanding and confusion, here we stress again that any $D(\theta,\phi)$ maps synthesized from the $d^D_{\lm}$ are {\it by no means} the CMB signals (since the phases of the these signals are not the phases of true CMB) and the true CMB can be obtained after multi-frequency analysis, which is the subject of our forthcomming paper.

\section{Acknowledgments}
We thank H.K. Eriksen, F.K. Hansen, A.J. Banday, C. Lawrence, K.M. Gorski and P.B. Lilje for their comments and critical remarks. We
acknowledge the use of NASA Legacy Archive for Microwave Background
Data Analysis (LAMBDA) and the maps. We also
acknowledge the use of the \healpix \citep{healpix} and the \glesp package \citep{glesp} to
produce $\alm$ from the \wmap data sets.

\appendix
\section{}
In this Appendix we would like to describe general properties of the $4n$ periodicity of the Galactic signal, taking into account its symmetry. We adopt the following model of the signal, which seems to be general. Let define some area around Galactic plane $S=\sum_{j=1}^N s^j_{pix}=Ns_{pix}$, where $s^j_{pix}$ is the pixel area and index $j$ mark the location of the pixel. We assume for simplicity that all the pixels in the map are have the same area. In polar system of coordinates corresponding angles $\theta_j$ and $\phi_j$ mark the position of $j$-th pixel in the map. Let us define the amplitude of the signal per each pixel as $ T_j$  . Thus the map which corresponds to the Galactic signal  is now
\begin{equation}
\Delta T(\theta,\phi)=\sum_j T_j \delta(\cos\theta-\cos\theta_j)\delta(\phi-\phi_j).
\label{A1}
\end{equation}
Let assumes that Galaxy image is localized in $\theta$-direction as  $ \pi/2-\delta\le\theta_j\le \pi/2+\delta$ and it could be or could not be localized in $\phi$-direction. Additionally we will assume that signal per each pixel $T_j$ is the sum of Galactic foreground signal $T^f_j$ and CMB plus instrumental noise signal $T^c_j$.
Important to note that statistical properties of these two components are different as in terms of amplitudes, as in terms of pixel-pixel correlations $\langle T_j T_k\rangle$. Particularly, in the area $S$ we have $T^f_j\gg
T^c_j$, while outside $S$ we assume that $T^f_j\ll T^c_j$.
Using proposed model of the signal in the map we can obtain corresponding $a_{\l,m}$ coefficients of the spherical harmonic expansion

\begin{equation}
a_{\l,m}=\sqrt{\frac{2l+1}{4\pi}\frac{(\l-m)!}{(\l+m)!}}\sum_j T_j P_\l^m(\cos\theta_j)e^{-im\phi_j},
\label{A2}
\end{equation}
which can be represented as a sum of foreground $F_{\l,m}$ coefficients and the CMB plus noise coefficients
$c_{\l,m}$. In order to understand the nature of $4n$-periodicity of the Galactic foreground, let discuss the model when $c_{\l,m}=0$.Then, from Eq.(\ref{eq1}) the subject of interest would be the phases of foreground
$\Phi_{\l,m}$ related to the $F_{\l,m}$ coefficients as follows
\begin{equation}
\tan\Phi_{\l,m}=-\frac{\sum_j T_j P_\l^m(\cos\theta_j)\sin(m\phi_j)}{\sum_j T_j P_\l^m(\cos\theta_j)\cos(m\phi_j)},
\label{A3}
\end{equation}
where sum over $j$ corresponds to the pixel in the area $S$.
Let's define the difference of phases, using their tangents.
\begin{equation}
\tan\Phi_{\l+\Delta,m}-\tan\Phi_{\l,m}=\frac{N}{D},
\label{A4}
\end{equation}
where
\begin{eqnarray}
N&=&\sum_{j,k}T_jT_k (P_\l^m(\cos\theta_j)P_{\l+\Delta}^m(\cos\theta_k)
\sin m\phi_j\cos m\phi_k -P_\l^m(\cos\theta_k)P_{\l+\Delta}^m(\cos\theta_j)\sin m\phi_k\cos m\phi_j;\nonumber\\
D&=&\sum_{j,k}T_jT_k P_\l^m(\cos\theta_j)P_{\l+\Delta}^m(\cos\theta_k)\cos m\phi_j\cos m\phi_k. \nonumber\\
\label{A5}
\end{eqnarray}
As one can see from Eq.(\ref{A4}), if $\Phi_{\l,m}\simeq \Phi_{\l+\Delta,m}$, then
$\tan\Phi_{\l,m}-\tan \Phi_{\l+\Delta,m}\simeq \Phi_{\l,m}- \Phi_{\l+\Delta,m}$ , which determine the properties of $d^{\Delta}_{\l,m}$ estimator for correlated phases (see Eq.(\ref{eq1}).
Below the object of out investigation is function $N$ from Eq.(\ref{A5}). Particularly we are interesting in
asymptotic $N\rightarrow 0$, which should reflect directly the symmetry of the foreground signal. Simple algebra allows us to represent $N$ function in the following form
\begin{equation}
N=\sum_{j,k}T_j T_k (P_\l^m(\cos\theta_j)P_{\l+\Delta}^m(\cos\theta_k)
-P_\l^m(\cos\theta_k)P_{\l+\Delta}^m(\cos\theta_j))\sin m(\phi_k-\phi_j).
\label{A6}
\end{equation}
Taking into account that area $S$ is located close to the $\theta=\pi/2$, let us discuss the properties of N-function at the limit $\l\pi/2\gg1$, using asymptotic of the Legendre polynomials. After
simple algebra we obtain
\begin{equation}
N\simeq \frac{1}{\pi}\sum_{j,k}\frac{T_jT_k}{\sqrt{\sin\theta_j\sin\theta_k}}
\frac{(\l+m)!(\l+\Delta+m)!}{\Gamma(\l+\frac{3}{2})\Gamma(\l+\Delta+\frac{3}{2})} G^m_{\l,\Delta}(\theta_j,\theta_k)\sin m(\phi_k-\phi_j),
\label{A7}
\end{equation}
where

\begin{eqnarray}
2G^m_{\l,\Delta}(\theta_j,\theta_k)=
\{\cos[(\l+\frac{1}{2})(\theta_j+\theta_k)+m\pi-\frac{\pi}{2}] +\cos[(\l+\frac{1}{2})(\theta_j-\theta_k)]\}(\cos\theta_k\Delta-\cos\theta_j\Delta) \nonumber \\
-\{\sin[(\l+\frac{1}{2})(\theta_j+\theta_k)+m\pi-\frac{\pi}{2}] -\sin[(\l+\frac{1}{2})(\theta_j-\theta_k)]\} (\sin\theta_k\Delta-\sin\theta_j\Delta). 
\label{A8}
\end{eqnarray}
From Eq.(\ref{A8}) one can see the symmetry of the Legendre polynomials which manifest themselfs trough $\cos\theta_j\Delta$ and $\sin\theta_j\Delta$ modes. If $\theta_j=\pi/2$, then depending on $\Delta$ we will have
\begin{eqnarray}
\cos(\frac{\pi}{2}\Delta)=1, \hspace{0.5cm} \Delta=4n, n=1,2...;\nonumber\\
\sin(\frac{\pi}{2}\Delta)=0, \hspace{0.5cm} \Delta=2n, n=1,2....\nonumber\\
\label{A9}
\end{eqnarray}
Thus, choosing $\Delta=4n, n=1,2...$ mode we take the corresponding properties of the Legendre polynomials into consideration. However, as one can see from Eq.(\ref{A9}) $\Delta=4n, n=1,2...$ periodicity of the Galactic image is not exact. In reality we have pixels which containt galaxy signal having $theta_j$ coordinate close to $\pi/2$, but not exactly equivalent to $\pi/2$. Let us introduce a new variable $\delta_j= \pi/2-\theta_j, \delta_j\ll 1$ characterized the deviation of the $j$-th
pixel location from the $theta=\pi/2$ plane. From Eq.(\ref{A8}) one can find that for $\Delta=4n, n=1,2...$
$\cos\theta_j\Delta\simeq \cos4n\delta_j\simeq 1-(4n\delta_j)2/2$. Thus , if pixel $j$ containt the signal from the galactic foreground, the deviation from the center of the galactic plane should be small enough:$\delta_j\ll \pi/4n$.
It is clear that this condition does not necessarily correspond to the properties of the Galactic image, which is clearly seen from the K, Ka and Q band signals.
Taking the above-mentioned properties of $G^m_{\l,\Delta=4n}(\theta_j,\theta_k)$ function, we represent the asymptotic of this
function at the limit $\delta_j\ll \pi/4n$, which is applicable for analysis of the Galactic signal at V and W bands.
\begin{eqnarray}
2G^m_{\l,\Delta=4n}(\theta_j,\theta_k)=
\frac{\Delta^2(\delta^2_j-\delta^2_k)}{2}\left\{\cos[(\l+\frac{1}{2})(\delta_k-\delta_j)]+(-1)^{\l+m} \cos[(\l+\frac{1}{2})(\delta_k+\delta_j)]\right\} \nonumber \\
+\Delta(\delta_j-\delta_k) \left\{\sin[(\l+\frac{1}{2})(\delta_k-\delta_j)]+(-1)^{\l+m} \sin[(\l+\frac{1}{2})(\delta_k+\delta_j)]\right\}. 
\label{A10}
\end{eqnarray}
Thus, combining Eq.(\ref{A7}) and Eq.(\ref{A10}), we obtain
\begin{eqnarray}
N\propto \sum_{j,k}T_jT_kG^m_{\l,\Delta=4n}(\theta_j,\theta_k)\sin m(\phi_k-\phi_j).
\label{A11}
\end{eqnarray}

One may think that choose of described above $4n$-mode of Legendre polynomials automatically   guarantee cancellation of the brightest part of the signal from the map without any restriction on symmetry and amplitude of the foreground.
To show that symmetry of the Galactic signal is important, let us discuss a few particular cases, which  illuminate this problem more clearly.

Firstly, let take a look at galactic center (GC), which is one of the brightest sources of the signal.
For the GC corresponding amplitudes $T_j$ are localized per pixels, for which $\phi_j\simeq 0$ in the galactic system of coordinate. From Eq.(\ref{A11}) one can see that  for GC the function $N$ is equivalent to zero . More accurately, taking into account that image of the GC has characteristic sizes $\delta\theta=\delta\phi\sim {\rm FWHM}$, where FWHM is the Full Width of Half Maximum of the beam, in Eq.(\ref{A11}) additionally to $\delta_j\ll 1$ parameter we get small parameter
$m{\rm FWHM}\ll 1$.

Secondly, let discuss the model of two bright point like sources, located symmetrically relatively to the GC. Let assumes that for that point sources $T_1\neq T_2$, but $\phi_2-\phi_1=\pi$. Once again, from Eq.(\ref{A11}) we get $N=0$ for all $m$ and these point sources will be automatically removed by the $d^{4n}_{\l,m}$ estimator even if they
have $\delta_1\neq\delta_2$.

Another possibility related to the symmetry of the Galactic image in $\theta$ direction. We would like to remind, that Eq.(\ref{A11}) was obtained under approximation $\l\theta_j\ll 1$, where $\theta_j=\pi/2-\delta_j$. This means, that
$\l\delta_j$ can be as big enough ($\l \delta_j \gg 1$), as small ($\l\delta_j\ll1$) as well. For $\l\delta_j\ll 1$ from
Eq.(\ref{A10}) we obtain 
\begin{equation}
2G^m_{\l,\Delta=4n}(\theta_j,\theta_k)=\frac{\Delta^2(\delta^2_j-\delta^2_k)}{2}[1+(-1)^{\l+m}]
+\Delta (\l+\frac{1}{2})[-(\delta_j-\delta_k)^2 +(-1)^{\l+m}(\delta^2_j-\delta^2_k)].
\label{A12}
\end{equation}
As one can see from Eq.(\ref{A12}) the bright sources located on the same $\theta$ coordinates ($\delta_j=\delta_k$)
does not contribute to $d^{4n}_{\l,m}$ estimator.

\end{document}